\documentclass{article}
\usepackage[utf8]{inputenc}
\usepackage{graphicx,anysize,amsfonts,amsmath,amsthm,amssymb,enumerate,hyperref,color,setspace,booktabs,enumitem,multirow,longtable,caption, bm, subcaption,bm,comment, setspace}
\usepackage[authoryear,numbers]{natbib}
\usepackage{amsmath, threeparttable, multirow, graphicx, array, bm, longtable, natbib}

\title{Analyzing  Clustered Continuous Response Variables with Ordinal Regression Models}

\author{Yuqi Tian, Bryan E. Shepherd, Chun Li, Donglin Zeng, Jonathan J. Schildcrout
}

\begin{document}

\date{}

\maketitle

\begin{abstract}
Continuous response variables often need to be transformed to meet regression modeling assumptions; however, finding the optimal transformation is challenging and results may vary with the choice of transformation.
When a continuous response variable is measured repeatedly for a subject or the continuous responses arise from clusters, it is more challenging to model the continuous response data due to correlation within clusters. We extend a widely used ordinal regression model, the cumulative probability model (CPM), to fit clustered continuous response variables based on generalized estimating equation (GEE) methods for ordinal responses. With our approach, estimates of marginal parameters, cumulative distribution functions (CDFs), expectations, and quantiles conditional on covariates can be obtained without pre-transformation of the potentially skewed continuous response data. Computational challenges arise with large numbers of distinct values of the continuous response variable, and we propose two feasible and  computationally efficient approaches to fit CPMs for clustered continuous response variables with different working correlation structures. We study finite sample operating characteristics of the estimators via simulation, and illustrate their implementation with two data examples. One studies predictors of CD4:CD8 ratios in an HIV study. The other uses data from The Lung Health Study to investigate the contribution of a single nucleotide polymorphism to lung function decline.

\textbf{Key words}: Clustered data; Cumulative probability model; Generalized estimating equation; Longitudinal data; Ordinal regression model.

\end{abstract}

\section{Introduction}
\label{sec:intro}

% challenges
Analyses of quantitative response variables are often challenged by distributions that do not follow standard parametric assumptions.  While it is common in such settings to transform the response variables so that model assumptions are satisfied, such transformations are often ad hoc and  parameters associated with the models can be difficult to interpret on their natural, untransformed scale.  For example, several studies of people living with HIV model associations with CD4:CD8 ratio, a biomarker that measures the strength of an individual's immune system. CD4:CD8 ratio tends to be right-skewed, and there is no standard accepted transformation. Researchers have analyzed CD4:CD8 ratio with no transformation \citep{castilho2016cd4}, log-transformation \citep{sauter2016cd4}, square-root transformation \citep{da2018impact}, fifth-root transformation \citep{gras2019determinants}, and various categorizations \citep{petoumenos2017cd4, serrano2017effects}. Finding the appropriate transformation can be challenging and results may be sensitive to the choice of transformation.

% take a step back - cross sectional continuous
A compelling approach to tackle the challenges associated with non-standard response distribution modeling is to treat continuous response variables as if they were ordinal using cumulative probability models (CPMs), also known as cumulative link models \citep{liu2017modeling}. The CPM is a semi-parametric linear transformation model \citep{zeng2006efficient} that assumes a linear model following an unspecified response transformation.  Rather than making an assumption about the appropriate transformation to apply, CPM fitting uses the data to estimate the transformation nonparametrically with a step function. The CPM is invariant to any monotonic transformation of the response variable because only order information is used for regression parameter estimation. Therefore, no pre-transformation of the response variable is needed.  Regression parameters from CPMs are interpretable, and because the cumulative distribution function (CDF) is modeled, conditional (on covariates) means and quantiles can be extracted from the CPM fit. The use of CPMs for cross-sectional continuous response variables, even with thousands of unique outcomes, is computationally feasible with applications of sparse matrix calculations and it has been implemented in Harrell's \texttt{orm()} function in the \textbf{rms} \textsf{R} package \citep{rms}. 

% our work (differences/extra work) 
Clustered continuous data are common in practice and important for studying exposure-response associations over time. The generalized estimating equation (GEE) procedure proposed by \citet{liang1986longitudinal} and \citet{zeger1986longitudinal} extends quasi-likelihood estimation \citep{wedderburn1974quasi} for generalized linear models (GLMs) \citep{mccullagh1983generalized}, from independent  to correlated data settings. Even though valid inferences are possible with GEE when second and higher 
order moments are misspecified, GEE for correlated data is challenged by non-standard distributions in the same way linear regression is for cross-sectional response data. Inspired by \citet{liu2017modeling}, in this paper, we present CPMs for clustered continuous response variables  to avoid specifying a transformation.
Specifically, we demonstrate that 1) CPMs can be fit to quantitative correlated data using GEE methods for ordinal data, and 2) GEE for ordinal data can be applied to non-standard, quantitative response distributions. Our proposed approach estimates time- and covariate-dependent CDFs, from which estimates of the mean, quantiles, and exceedance probabilities can be derived. In addition, we present software and strategies for implementing GEE methods for ordinal data to settings with large numbers (i.e., hundreds or thousands) of distinct levels. 

% outline of paper
In Section 2, we review CPMs for cross-sectional continuous response variables. In Section 3, we demonstrate how CPMs for clustered data can be fit using GEE for ordinal response variables, and we propose practical estimation techniques. We illustrate the performance of the methods by simulation in Section 4. In Section 5, we apply our methods to data from two studies.  The first investigates predictors of CD4:CD8 ratio in a longitudinal cohort of people living with HIV. The second evaluates the genetic contribution of a single nucleotide polymorphism to lung function decline in a cohort of smokers with chronic obstructive pulmonary disease (COPD).  Finally, we discuss strengths and limitations of the proposed methods and potential future directions in Section 6.

\section{Review of Methods}\label{sec:review}
\label{s:model}

The CPM is a class of models for scalar ordinal response data \citep{liu2017modeling}. 
Let $Y$ be a continuous response variable, and $Y^* = h(Y)$ be a transformation of $Y$ with $h(\cdot)$ an unspecified non-decreasing function. Let $\bm{X}$ be a vector of covariates with $\bm{X} = \bm{0}$ as a reference value. Let $\epsilon$ be an error term. We assume the relationship between the transformed variable and covariates is linear, $Y^*=\bm{\beta}^T\bm{X}+\epsilon$, where $\epsilon$ follows a known distribution $F_\epsilon$ and $\bm{\beta}$ is a vector of regression parameters. It follows that
\begin{align}
    Y = h^{-1}(Y^*)=h^{-1}(\bm{\beta}^T \bm{X}+\epsilon).
    \label{eq:linear_trans}
\end{align}
Letting $G=F_\epsilon^{-1}$ be a link function. (\ref{eq:linear_trans}) can be expressed as a CPM with
\begin{align*}
\begin{gathered}
    F(y|\bm{X})=P(Y\le y|\bm{X})=P\left(\epsilon\le  h(y)-\bm{\beta}^T \bm{X}|\bm{X}\right) =F_\epsilon\left(h(y)-\bm{\beta}^T \bm{X}\right), \text{ which implies}\\
    G\{F(y|\bm{X})\}=h(y)-\bm{\beta}^T \bm{X}.
\end{gathered}
\end{align*}
The intercept $h(y)=G\{F(y|\bm{X}=\bm{0})\}$ represents the link-transformed CDF for $\bm{X}=\bm{0}$ (i.e. the reference CDF), and $\bm{\beta}^T\bm{X}$ represents shifts in this CDF that depend on the values of $\bm{X}$. The interpretation of $\bm{\beta}$ depends on the choice of the link function/$F_\epsilon$. For example, $\bm{\beta}$ is interpreted as a log odds ratio with the logit link (i.e., $F_{\epsilon}$ logistic distribution) and a log hazard ratio with the complementary log-log link (i.e., $F_{\epsilon}$ extreme value distribution).

Assume there are $N$ subjects and denote $y_{(j)}$ to be the $j$th smallest observed response value ($j=1, \dots, J$). Rather than specifying a functional form for $h(\cdot)$, we can estimate it using a step function with $\gamma_j=h(y_{(j)})$. Since $h(\cdot)$ is estimated nonparametrically in the CPM, it belongs to the class of semi-parametric linear transformation models \citep{zeng2006efficient, zeng2007maximum}.  For $(y_i, \bm{X}_i), i\in 1, \dots, N$, the CPM is given by
\begin{align}
    G\{F(y_i|\bm{X}_i)\}= G\{F(y_{(j)}|\bm{X}_i)\}= \gamma_j-\bm{\beta}^T \bm{X}_i.
    \label{eq:cpm_iid}
\end{align}
Letting $\bm{\theta}=(\bm{\gamma}^T, \bm{\beta}^T)^T$ and $\bm{\gamma}=(\gamma_1, \dots, \gamma_{J-1})^T$, the likelihood is
\begin{align}
    L(\bm{\theta})=\prod_{j=1}^J\prod_{i:y_i=y_{(j)}}\left\{F
(y_i|\bm{X}_i)-F(y_i^-|\bm{X}_i)\right\},
\label{eq:first_l}
\end{align}
where $F(y_i^-|\bm{X}_i)=\lim_{t \uparrow y_i}F(t|\bm{X}_i)$. A nonparametric likelihood can be obtained by substituting $F(y_{(j-1)}|\bm{X}_i)$ for $F(y_{(j)}^-|\bm{X}_i)$  as follows
\begin{align}
    L(\bm{\theta})=\prod_{j=1}^J\prod_{i:y_i=y_{(j)}}\left\{G^{-1}(\gamma_j-\bm{\beta}^T \bm{X}_i)-G^{-1}(\gamma_{j-1}-\bm{\beta}^T \bm{X}_i)\right\},
    \label{eq:final_l}
\end{align}
where $-\infty \equiv \gamma_0 < \gamma_1 < \dots < \gamma_{J-1} < \gamma_J \equiv \infty$. From this likelihood, nonparametric maximum likelihood estimates (NPMLEs) of $\bm{\theta}$ can be obtained.

The CPM in (\ref{eq:cpm_iid}) is identical to the cumulative link model used for ordinal data. For example, the CPM with the logit link is referred to as the proportional odds model. The likelihood in (\ref{eq:final_l}) is identical to the multinomial likelihood used to estimate parameters of cumulative link models for ordinal data \citep{snell1964scaling, mccullagh1980regression, agresti2010analysis}. It follows that a semi-parametric linear transformation model can be fit using an ordinal CPM where each distinct value of continuous $Y$ is treated as its own ordinal category. With truly continuous $Y$, there will be $N$ such categories. To summarize briefly, with CPMs, a continuous response variable CDF is modeled as a linear function of covariates after an unspecified monotonic transformation is applied. The transformation is estimated nonparametrically from the observed data with a step function. 

CPMs have a number of attractive properties for fitting continuous response data \citep{liu2017modeling, tian2020empirical}. First, since only ordinal information is incorporated for estimating $\bm{\beta}$, CPMs are invariant to any monotonic transformation of response variables, which means no transformation of response variables is needed. They also work well with continuous response variables subject to detection limits even with high censoring rates and small sample sizes \citep{tian2022detection}.  It has been shown that under some mild conditions, CPMs result in estimates that are consistent and asymptotically normal \citep{li2022asymptotics}, and their variances can be estimated with the inverse of the observed information matrix.  
Other quantities, such as quantiles, exceedance probabilities,  and expectations conditional on covariates can be derived from the CPM model fit. For example, the expectation can be estimated with $\hat{E}(Y|\bm{X})=\sum_{j=1}^J\sum_{i:y_i=y_{(j)}} y_{(j)} \left\{ \hat{F}(y_{(j)}|\bm{X}) - \hat{F}(y_{(j-1)}|\bm{X})\right\}$. Standard errors for CDFs and expectations can be calculated using the delta method \citep{liu2017modeling}, and quantiles can be estimated with linear interpolations of the inverse of the CDFs \citep{liu2017modeling, tian2022detection}. 

Until recently, the use of CPMs for continuous responses was rare due, in part, to computational costs. Harrell's \texttt{orm()} function in the \textbf{rms} package in \textsf{R} is a computationally efficient implementation of CPMs that can be fit with tens of thousands of distinct responses. The \texttt{orm()} function takes advantage of the sparse structure of the Hessian matrix which allows for efficient inversion by Cholesky decomposition in a Newton-Raphson algorithm \citep{rms, liu2017modeling}.

\section{Methods}

\subsection{CPMs for Clustered Continuous Response Variables}

We extend CPMs to the cluster correlated response setting for the same reason they were developed in the cross-sectional response setting; namely, we would like to avoid parametric and often ad hoc transformations of the response to satisfy modeling assumptions.

% notation
Suppose there are $N$ subjects, $i\in \{1, \dots, N\}$ indexes subjects, and subject $i$ has $T_i$ observations. Denote the response for subject $i$ at time $t$ with $Y_{it}$, and $\bm{Y}_i = (Y_{i1}, \dots, Y_{iT_i})^T$. Across all subjects, $\bm{Y}=(\bm{Y}_1, \dots, \bm{Y}_N)^T$ has a total of $J$ distinct values; with truly continuous $\bm{Y}$, $J=\sum_{i=1}^N T_i$.  Let $Z_{it,j}=I(Y_{it}\le y_{(j)})$ and $\mu_{it,j}=E(Z_{it,j}|\bm{X}_{it})=P(Y_{it}\le y_{(j)}|\bm{X}_{it})=F(y_{(j)}|\bm{X}_{it})$, where $y_{(j)}$ corresponds to the $j$th smallest value among the $J$ levels of the response variable, and $\bm{X}_{it}$ is  the design vector for subject $i$ at time $t$. Let the vector of binary indicator variables for subject $i$ at time $t$ be $\bm{Z}_{it}=(Z_{it,1}, \dots, Z_{it,J-1})^T$, and $\bm{\mu}_{it}=(\mu_{it,1}, \dots, \mu_{it,J-1})^T$. Finally, for subject $i$, let $\bm{Z}_i=(\bm{Z}_{i1}^T, \dots, \bm{Z}_{iT_i}^T)^T$ and $\bm{\mu}_i=(\bm{\mu}_{i1}^T, \dots, \bm{\mu}_{iT_i}^T)^T$.

Suppose $Y_{it}$ has a linear relationship with the covariates $\bm{x}_{it}$ after an unspecified monotonic transformation $h(\cdot)$.  This leads to a linear transformation model
\begin{align}
    Y_{it}=h^{-1}(Y_{it}^*)=h^{-1}(\bm{\beta}^T \bm{X}_{it} + \epsilon_{it}),
    \label{eq:linear_transformation_model}
\end{align}
where $\epsilon_{it}$ follows a specified distribution and $\epsilon_{it}$ is independent of $\epsilon_{i't'}$ for $i\neq i'$, but not necessarily independent if $i=i'$. Let $G=F_{\epsilon}^{-1}$ be a link function. Based on the linear transformation model, we have $\mu_{it,j}=P(Y_{it}\le y_{(j)}|\bm{X}_{it})
    %&=P(h^{-1}(\bm{\beta}^T \bm{x}_{it}+\epsilon_{it})\le y_{(j)}|\bm{x}_{it})\\
    =P(\epsilon_{it} \le  h(y_{(j)})-\bm{\beta}^T \bm{X}_{it}|\bm{X}_{it})
    =F_\epsilon(h(y_{(j)})-\bm{\beta}^T \bm{X}_{it})$, which implies 
    $G(\mu_{it,j}) = h(y_{(j)})-\bm{\beta}^T \bm{X}_{it}$.
Therefore, similar to (\ref{eq:cpm_iid}), the CPM for a clustered continuous response variable is:
\begin{align}
    G(\mu_{it,j})=\gamma_j - \bm{\beta}^T \bm{X}_{it},
    \label{eq:marginal_model_continuous}
\end{align}
where $G(\cdot)$ is the specified link function, $\gamma_j=h(y_{(j)})$, and $\bm{\theta}=(\bm{\gamma}^T, \bm{\beta}^T)^T$. Like all models, the interpretation of $\bm{\beta}$ depends on the link function. For example, if $G(\cdot)$ is the log odds link, $\bm{\beta}$ is a log odds ratio; if $G(\cdot)$ is the log-log link, $\bm{\beta}$ is a log hazard ratio. The intercepts $\bm{\gamma}$ are the link function transformed CDFs when  all covariates set equal to 0. This also represents the transformation needed for the response variable to be modeled by a linear model. 

With clustered data, we cannot directly apply nonparametric maximum likelihood estimation to fit CPMs because observations are not independent. Since the CPM is parameterized as an expectation $\mu_{it,j}=E(Z_{it,j}|\bm{X}_{it})$, we can rely on GEE techniques to estimate parameters in (\ref{eq:marginal_model_continuous}).  To obtain valid inferences, GEE requires correct specification of the marginal model for the response mean. GEE also permits specification of within cluster response dependence with a working correlation structure. The working correlation structure does not have to represent the true structure; however, to the extent that it differs from the true structure, efficiency losses incur \citep{liang1986longitudinal, zeger1986longitudinal}. 
GEE methods for longitudinal ordinal responses have been discussed in a number of papers \citep{ heagerty1996marginal, lipsitz1994analysis, huang2002building, parsons2006generalized, touloumis2013gee}.

We estimate $\bm{\theta}$ in (\ref{eq:marginal_model_continuous}) using GEE methods for ordinal response data by solving the estimating equation
\begin{align}
% A or B
     A_{\bm{\theta}}(\bm{\theta};\bm{\alpha}) = \sum_{i=1}^N \bm{D}_i^T \bm{W}_i^{-1} (\bm{Z}_i-\bm{\mu}_i) = \bm{0},
    \label{eq:cpm_gee1}
\end{align}
where $\bm{D}_i=\frac{\partial \bm{\mu}_i}{\partial \bm{\theta}}$, $\bm{W}_i=\bm{S}_i^{\frac{1}{2}} \bm{R}_i(\bm{\alpha}) \bm{S}_i^{\frac{1}{2}}$, and $\bm{\alpha}$ is a vector of association parameters. $\bm{R}_i(\bm{\alpha})$ is a working correlation matrix for $\bm{Z}_i$ and $\bm{S}_i$ is a $T_i(J-1)\times T_i(J-1)$ block matrix with elements based on the variance of $Z_{it,j}$, $\{\mu_{it,j} ( 1-\mu_{it,j})\}^{\frac{1}{2}}$. $\bm{W}_i^{-1}$ can be considered as a weight matrix for subject $i$. Efficiency is improved to the extent that the working correlation matrix $\bm{R}_i(\bm{\alpha})$ is a better approximation to the true correlation structure of $\bm{Z}_i$. The structure of $\bm{R}_i(\bm{\alpha})$ is assumed by the analyst and $\bm{\alpha}$ can then be estimated with a second estimating function that will be described in more detail in Section \ref{sec:cpm_uni}.
 
The covariance of $\bm{\theta}$ is given by
\begin{align}
    V_{\bm{\theta}}(\bm{\alpha})=\left(\sum_{i=1}^N \bm{D}_i^T \bm{W}_i^{-1} \bm{D}_i\right)^{-1} \left(\sum_{i=1}^N \bm{D}_i^T \bm{W}_i^{-1} \text{Cov}(\bm{Z}_i) \bm{W}_i^{-1} \bm{D}_i\right)
    \left(\sum_{i=1}^N \bm{D}_i^T \bm{W}_i^{-1} \bm{D}_i\right)^{-1},
    \label{eq:asym_cpm_gee_cov}
\end{align}
which can estimated by replacing $\bm{\theta}$ with $\hat{\bm{\theta}}$ and $\text{Cov}(\bm{Z}_i)$ with $(\bm{Z}_i-\bm{\mu}_i)(\bm{Z}_i-\bm{\mu}_i)^T$.

Since $\mu_{it,j}=F(y_{(j)}|\bm{x}_{it})$ is a CDF, other quantities can be readily obtained from a fitted CPM. The CDF can be calculated with $\hat{F}(y|\bm{X})=G^{-1}(\hat{\gamma}_j - \hat{\bm{\beta}}^T \bm{X})$, where $j$ is the index such that $y_{(j)} = \max\{ j' \in \{1, \dots, J\}: y_{(j')} \le y\}$. We can derive its standard error with the delta method. %The estimated CDFs can be useful if probabilities of the response variable below or above certain thresholds are of interest in practice. 
Similar to the scalar response setting, cross-sectional summaries (e.g. quantiles, exceedance probabilities, and expectations) an can calculated from $\hat{\bm{\theta}}$ and $\hat{V_{\bm{\theta}}}(\bm{\alpha})$.

% Challenges fitting this model + preface the next 2 subsections
It is worth noting that fitting ordinal GEE methods to clustered continuous response data is computationally challenging. Specifically, for each observation $Y_{it}$, we need $J-1$ indicators $Z_{it,j}=I(Y_{it}\le y_{(j)})$, and $J$ is usually a large number for continuous data, which implies that $\bm{W}_i$ and $\bm{D}_i$ in (\ref{eq:cpm_gee1}) and (\ref{eq:asym_cpm_gee_cov}) can be high-dimensional. In the following subsections, we will introduce two feasible and computationally efficient implementations to analyze clustered continuous response variables based on CPMs. We first consider the relatively straightforward case with independence working correlation structures. We then move on to more complex working correlation structures that are commonly implemented for GEE-based estimation.

\subsection{CPMs with Independence Working Correlation}

It is well known that working covariance weighting can be more efficient than working independence weighting, particularly for parameters corresponding to time-varying covariates. However, the independence working correlation structure is simpler and therefore easier to implement than other structures because it does not require estimating $\bm{\alpha}$, and the computation burden of matrix inversion is reduced with a diagonal structure. In addition, there are settings where using an independence working correlation structure is recommended for statistical reasons, the most common of which occurs when interest is in the cross-sectional $E(Y_{it}|\bm{X}_{it})$ but where $E(Y_{it}|\bm{X}_{it}) \neq E(Y_{it}|\bm{X}_{i1}, \dots, \bm{X}_{iT_i})$. In such settings, one must use an independence working correlation to ensure consistent estimates of time-varying covariate parameters \citep{pepe1994cautionary, schildcrout2005regression, diggle2002analysis}. There are many examples in practice where the cross-sectional conditional expectation may be of interest but is not equal to the full conditional expectation (e.g., \citet{lauderdale2008sleep}).

As described in Section 2, CPMs for scalar response data can be fit to response data with thousands of distinct values.  With an independence working correlation structure, solving (\ref{eq:cpm_gee1}) for $\bm{\theta}$ and plugging $\hat{\bm{\theta}}$ into (\ref{eq:asym_cpm_gee_cov}) to estimate the variance is equivalent to treating the response data as unclustered, computing the NPMLEs of CPMs as described in Section 2, and then correcting estimates of uncertainty by using a sandwich-variance estimate (see Web Appendix A in Supporting Information). Therefore, CPMs with independence working correlation can be expeditiously fit to clustered continuous responses with thousands of distinct values. We fit CPMs to clustered continuous response variables by maximizing the marginal likelihood
\begin{align}
\begin{split}
    L(\bm{\theta}) &=\prod_{j=1}^J\prod_{i,t:y_{it}=y_{(j)}}\left(F
(y_{it}|\bm{X}_{it})-F(y_{it}^-|\bm{X}_{it})\right)\\ 
    &=\prod_{j=1}^J\prod_{i,t:y_{it}=y_{(j)}}\left(G^{-1}(\gamma_j-\bm{\beta}^T \bm{X}_{it})-G^{-1}(\gamma_{j-1}-\bm{\beta}^T \bm{X}_{it})\right)\\
    &=\prod_{j=1}^J\prod_{i,t:y_{it}=y_{(j)}}\left(\mu_{it,j} - \mu_{it,j-1}\right).
    \label{eq:longitudinal_l}
\end{split}
\end{align}
To correct for correlated responses within each cluster, we use the Huber sandwich estimator to estimate the covariance \citep{huber1967under, white1980heteroskedasticity, freedman2006so}. Since the clusters are independent but observations are dependent, we group observations within clusters. 
Let 
\begin{align*}
    l(\bm{\theta})=\log\left(L(\bm{\theta})\right)=\sum_{j=1}^J\sum_{i,t:y_{it}=y_{(j)}} \log\left( f_{it,j} \right)
\end{align*} be the log-likelihood of (\ref{eq:longitudinal_l}) under the assumption of independent observations, where $f_{it
,j}= \mu_{it,j}-\mu_{it,j-1}$. 
The first and second order partial derivatives of $l(\bm{\theta})$ with respect to $\bm{\theta}$ are given by
\begin{align*}
    l'(\bm{\theta})&=\frac{\partial l(\bm{\theta})}{\partial \bm{\theta}}=\sum_{j=1}^J\sum_{i,t:y_{it}=y_{(j)}}  \frac{\partial \log\left(f_{it,j}\right)}{\partial \bm{\theta}} = \sum_{j=1}^J\sum_{i,t:y_{it}=y_{(j)}} g_{it,j},\\
    l''(\bm{\theta})&=\frac{\partial^2 l(\bm{\theta})}{\partial \bm{\theta}^2}=\sum_{j=1}^J\sum_{i,t:y_{it}=y_{(j)}}  \frac{\partial^2  \log\left(f_{it,j}\right)}{\partial \bm{\theta}^2},
\end{align*}
%Let $c_i$ be the observations of subject $i$. 
and Huber-White sandwich estimator for $\text{Cov}(\hat{\bm{\theta}})$ is given by
\begin{align}
    \left( l''(\hat{\bm{\theta}}) \right)^{-1}
    \left(\sum_{i=1}^N \left(\sum_{t=1}^{T_i}  \hat{g}_{it,j} \right) \left(\sum_{t=1}^{T_i}  \hat{g}_{it,j} \right)^T \right)
    \left( l''(\hat{\bm{\theta}}) \right)^{-1},
\label{eq:sandwich}
\end{align}
where $\sum_{t=1}^{T_i}  \hat{g}_{it,j}$ is the sum of the plug-in estimators for the first partial derivative elements within a cluster. Consistency and asymptotic normality of estimates and the validity of the sandwich estimators using this approach are shown under the conditions provided in Web Appendix B in Supporting Information \citep{li2022asymptotics}. 
Point estimates and robust covariances for CPMs can be obtained by the \texttt{orm()} and \texttt{robcov()} functions in the \textbf{rms} package in \textsf{R}, respectively \citep{rms}. 

\subsection{CPMs with Exchangeable/AR1 Working Correlation}\label{sec:cpm_uni}

Though computationally efficient, CPMs with independence working correlation structure can be statistically inefficient if the within cluster correlation is high and/or clusters are large. GEE methods for ordinal response variables  allow for more complicated working correlation structures to improve efficiency. \citet{lipsitz1994analysis} estimated association parameters with Pearson residuals; \cite{heagerty1996marginal} extended alternating logistic regression for binary longitudinal outcomes to ordinal longitudinal outcomes using pairwise log-odds ratio parameters as the association parameters \citep{lipsitz1991generalized, carey1993modelling}; \citet{touloumis2013gee} captured response association with local odds ratios based on Goodman's row and column effects models.

To improve efficiency over the independence working correlation approach described above, we appeal to the framework proposed by \citet{parsons2006generalized, parsons2009repeated} that specifies the association parameter $\bm{\alpha}$ as a correlation and that estimates the parameter iteratively by minimizing the determinant of $V_{\bm{\theta}}(\bm{\alpha})$. This method, which \citet{parsons2009repeated} called ``repolr'' (repeated measures proportional odds logistic regression), estimates $\bm{\alpha}$ based on the covariance matrix, whose dimension is manageable. In contrast, other ordinal GEE methods require enumerating all pairs of observations within each cluster to estimate $\bm{\alpha}$, which is extremely computationally intensive for continuous response data. In repolr, $\bm{R}_i(\bm{\alpha})$ is constructed as $\bm{R}_i(\bm{\alpha})=\bm{K}_i(\bm{\alpha}) \otimes \bm{C}$, where  $\bm{K}_i(\bm{\alpha})$ is a $T_i\times T_i$ within cluster working correlation matrix and $\bm{C}$ is a $(J-1)\times (J-1)$ matrix of correlations among  elements in $\bm{Z}_{it}$. By assumption, $\bm{C}$ is the same for every pair of binary indicators of ordinal levels for every subject at every time point, so that
$$\bm{C}=\left[\begin{matrix}\rho_{11} & \dots & \rho_{1(J-1)}\\\vdots & \ddots  & \vdots\\\rho_{(J-1)1} & \dots & \rho_{(J-1)(J-1)}\\\end{matrix}\right],$$
where  $\rho_{pq}$ is expected correlation between $Z_{itp}$ and $Z_{itq}$ for $i=1, \dots, N$. With the logit link, $\rho_{pq}=\rho_{qp}=\{\exp(\gamma_p - \gamma_q)\}^{\frac{1}{2}}$ where $p<q$ \citep{kenward1994application}.
Two common structures for $\bm{K}(\alpha)$ are exchangeable (also called uniform or compound symmetric) and first-order autoregressive (AR1) structures \citep{diggle2002analysis}, where only a single association parameter is used. For the exchangeable structure, $\bm{K}_{(p,q)}(\alpha)=1$ if $p=q$ and $\bm{K}_{(p,q)}(\alpha)=\alpha$ otherwise; for AR1 structure, $\bm{K}_{(p,q)}(\alpha)=1$ for $p=q$ and $
\bm{K}_{(p,q)}(\alpha)=\alpha^{|t_p - t_q|}$ otherwise. The additional estimating equation for the association parameter $\alpha$ in repolr is 
\begin{align}
    \frac{\partial \log |V_{\bm{\theta}}(\alpha)|}{\partial \alpha}=0,
    \label{eq:repolr2}
\end{align}
which is equivalent to estimating $\alpha$ by minimizing $\log |V_{\bm{\theta}}(\alpha)|$. That is, this equation solves for the  $\alpha$ that minimizes the confidence region size of the $\bm{\theta}$ parameter estimates. 
The algorithm iterates between solving (\ref{eq:cpm_gee1}) for $\hat{\bm{\theta}}$ and solving (\ref{eq:repolr2}) for $\hat{\alpha}$ until convergence. This approach can be applied with the \texttt{repolr()} function in the  \textbf{repolr} package in \textsf{R} \citep{parsons2017repolr} for complete data and for the logit link.

With continuous response variables, it may still be expensive to run a fully-iterated repolr model; hence, we propose a one-step GEE estimator for repolr \citep{lipsitz2017one}. In our setting, instead of iterating between the two estimating equations (\ref{eq:cpm_gee1}) and (\ref{eq:repolr2})  until convergence, we start with an estimate of $\bm{\theta}$ under an independence working correlation structure, $\hat{\bm{\theta}}_I$, which can be efficiently estimated with CPMs.  We then obtain the association parameter $\hat{\alpha}$ by solving (\ref{eq:repolr2}) with $V_{\hat{\bm{\theta}}_I}(\alpha)$. Finally, we solve (\ref{eq:cpm_gee1}) using $\hat{\alpha}$ to get $\hat{\bm{\theta}}$, which is asymptotically equivalent to the fully-iterated GEE estimator \citep{lipsitz2017one}.

% software
We built an \textsf{R} package, \textbf{cpmgee} (available at https://github.com/YuqiTian35/cpmgee), that applies this one-step estimation procedure for exchangeable and AR1 working correlation structures. This package also fits CPMs with independence working correlation. 

Although the one-step GEE estimator for repolr can substantially reduce the  computational burden, computation with exchangeable and AR1 working correlation structures may still be intensive if the number of distinct values of a continuous response variable is large. For this reason, one may seek to reduce the number of distinct values in the response by binning. Specifically, the $N'=\sum_{i=1}^N T_i$ observations can be divided into $M_b$ bins, where the value assigned to each observation in the bin is the median value for observations in that bin. Approximately equal-quantile binning can be achieved by expressing $N'$ as
\begin{align*}
    N'=M_b q + r = (M_b - r)q + r(q + 1),
\end{align*}
where $q$ is the integer quotient of $\frac{N'}{M_b}$. In this way, $M_b-r$ bins have $q$ observations, and $r$ bins have $q+1$ observations. Rounding is yet another way to reduce the number of distinct values.  More strategies for binning and rounding for cross-sectional CPMs with very large sample sizes are provided elsewhere \citep{li2022big}.

\section{Simulations}

% setting
We studied the performance of our estimators applying CPMs with independence, exchangeable, and AR1 working correlation to continuous clustered data under various simulation settings. Responses were generated in the following manner for subject $i$ at time $t$:
$$Y_{it}=\text{Inv-}\chi^2\left(\frac{\Phi(Y_{it}^*)}{2}, \text{ df}=5\right), \text{ and } Y_{it}^*=X_{i}\beta_X+ T_{it}\beta_T + \epsilon_{it},$$
where $\text{Inv-}\chi^2(\cdot, \text{ df=5})$ is the inverse of the CDF for a chi-square distribution with 5 degrees of freedom and $\Phi(\cdot)$ is the probability density function of the standard normal distribution. The transformation has been used in earlier work \citep{tian2020empirical} and was chosen because it does not correspond to a commonly-used closed-form transformation.

% primary setting
In the primary setting,  we set the sample size $N$ to be 1000, and imposed dropout completely at random uniformly from $t\in\{2, 3, 4, 5, 6\}$. $X_{i}$ was a time-invariant covariate following the standard normal distribution, $T_{it}$ represented time, a time-varying covariate, and was set to be $0, 0.2,\dots, 1$. A logistic residual distribution was used and the correlation structure was exchangeable with $\alpha=0.7$. We set $\beta_X=1$ and $\beta_T=1$. %In Figure \ref{fig:transformation_example}, we show histograms of the response variable after different transformation based on one simulated data set. The response variable on its original scale is right-skewed. A natural choice for right-skewed data is log-transformation. However, the log-transformed response variable is slightly left-skewed. The correct transformation $2\Phi^{-1}(\chi^2(\cdot, \text{ df=5}))$ is not a function one would typically consider for transformation.  
For CPMs with exchangeable and AR1 working correlation structures, we fit models using equal-quantile binning with $M_b=300$. 

%\begin{figure}
%    \centering
%    \includegraphics[scale=0.7]{plot/transformation_example.pdf}
%    \caption{(A) Histogram of the response variable. (B) Histogram of the %log-transformed response variable. (C) Histogram of the response %variable with the correct transformation $2\Phi^{-1}(\chi^2(Y, %\text{df}=5))$ for a linear model.}
%    \label{fig:transformation_example}
%\end{figure}

% other settings 
In addition to the primary setting, we also explored scenarios with a smaller $\alpha$, different values of $M_b$ for equal-quantile binning, and rounding with different decimal places. Additional simulation settings including the identity transformation (i.e., $Y=Y^*$); complete data; differing sample sizes, cluster sizes, time effects, and correlation structures; and link function misspecification are shown in Web Appendix C in Supporting Information.

% evaluation
We replicated each scenario 1000 times and evaluated operating characteristics with percent bias, root mean squared error (RMSE), empirical standard error, average estimated standard error, and coverage of 95\% confidence intervals. We also compared our CPM methods with standard GEE methods for continuous data with the correctly transformed response variable; which under the correct transformation and correlation structure, is optimal for estimating $\bm{\beta}$. 
We also investigated the performance of estimates of the conditional expectation, median, and CDF -- specifically, $E(Y|X=1,T=0.2)$, $Q(0.5|X=1,T=0.2)$, and $F(5|X=1, T=0.2)$, respectively -- that were estimated from the fitted CPMs. We do not show the average estimated standard error of $Q(0.5|X=1,T=0.2)$ because its confidence interval was obtained from linear interpolation of the inverse of the confidence interval for the conditional CDF.
% relative efficiency
%To evaluate the relative efficiency (RE) of regression parameters, we divided the empirical variance obtained from a CPM method by the empirical variance of standard GEE for continuous response variables with the correct transformation. 

Computation time for the CPM fits with independence and exchangeable working correlation is shown in Web Appendix D in Supporting Information. CPM fits with independence working correlation are very computationally efficient and can handle thousands of distinct values in the response variable.

\subsection{The Primary Setting}

\begin{table}
\caption{Simulation results for CPMs for the primary setting and its modifications with lower within cluster correlation ($\alpha=0.3$). For comparison, standard GEE models were fit with the correct transformation and the correct exchangeable working correlation structure.}
    \label{tab:primary}

    %\begin{tabular}{@{\extracolsep{5pt}}c c ccccc@{}}
    \begin{tabular}{c c c ccccc}
    \hline
    $\alpha$ & Method & Metric & $\beta_X$ & $\beta_T$ & ${\scriptstyle E(Y|X=1,T=0.2)}$   & ${\scriptstyle Q(0.5|X=1,T=0.2)}$ & ${\scriptstyle F(5|X=1,T=0.2)}$\\
    \hline
    \multirow{18}{*}{0.7}&  & Bias(\%) & -0.010 & 0.087 & - & - & -\\
    &   & RMSE & 0.050 & 0.060 & - & - & -\\
    & GEE & Empirical SE & 0.050 & 0.060 & - & - & -\\
    & (ex)  & Average SE & 0.051 & 0.059 & - & - & -\\
    &  & Coverage & 0.953 & 0.944 & - & - & -\\
    & & RE & \text{reference} & \text{reference} & - & - & - \\
    \cline{2-8}
    & & Bias(\%) & 0.129 & 0.270 & -0.009 & -0.074 & -0.169\\
      &  & RMSE & 0.054 & 0.091 & 1.232 & 1.199 & 0.171\\
    & CPM & Empirical SE & 0.054 & 0.091 & 0.139 & 0.132 & 0.016 \\
     & (ind)  & Average SE & 0.055 & 0.088 & 0.142 & - & 0.016\\
    &  & Coverage & 0.957 & 0.942 & 0.956 & 0.958 & 0.956\\
    & & RE & 1.129 & 2.279 & - & - & - \\
    \cline{2-8}
    & & Bias(\%) & 0.234 & 2.983 & -0.181 & -0.270 & -0.077\\
    & & RMSE & 0.052 & 0.075 & 1.224 & 1.191 & 0.170\\
    & CPM & Empirical SE & 0.052 & 0.069 & 0.135 & 0.130 & 0.015 \\
    & (ex)  & Average SE & 0.053 & 0.067 & 0.137 & - & 0.016\\
    &  & Coverage & 0.957 & 0.910 & 0.948 & 0.956 & 0.958\\
    & & RE & 1.047 & 1.310 & - & - & -\\
    \hline
    
    \multirow{18}{*}{0.3} &  & Bias(\%) & -0.061 & 0.127 & - & - & -\\
    & & RMSE & 0.040 & 0.089 & - & - & -\\
    & GEE & Empirical SE & 0.040 & 0.089 & - & - & -\\
    & (ex)  & Average SE & 0.040 & 0.087 & - & - & -\\
    & & Coverage & 0.955 & 0.943 & - & - & -\\
    & & RE & \text{reference} & \text{reference} & - & - & -\\
    \cline{2-8}
    & & Bias(\%) & 0.063 & 0.254 & -0.015 & -0.054 & -0.069\\
     & & RMSE & 0.041 & 0.092 & 1.236 & 1.204 & 0.171\\
    & CPM & Empirical SE & 0.041 & 0.092 & 0.107 & 0.105 & 0.013 \\
     & (ind)  & Average SE & 0.042 & 0.091 & 0.105 & - & 0.013\\
     & & Coverage & 0.959 & 0.946 & 0.957 & 0.952 & 0.959\\
    & & RE & 1.063 & 1.073 & - & - & - \\
    \cline{2-8}
    & & Bias(\%) & 0.160 & 2.929 & -0.196 & -0.249 & 0.023\\
    &  & RMSE & 0.041 & 0.091 & 1.227 & 1.195 & 0.170\\
    & CPM & Empirical SE & 0.041 & 0.086 & 0.105 & 0.105 & 0.013 \\
    & (ex)  & Average SE & 0.041 & 0.085 & 0.109 & - & 0.013\\
    & & Coverage & 0.961 & 0.936 & 0.953 & 0.943 & 0.959\\
    & & RE & 1.041 & 0.943 & - & - & -\\
    \hline
    \end{tabular}
\end{table}

% primary setting
Simulation results under the primary setting with $\alpha=0.7$ and modification with $\alpha=0.3$ are shown in Table \ref{tab:primary}. For the primary setting $(\alpha=0.7)$, CPMs performed quite well with low bias and generally good coverage for $\beta_X$, $\beta_T$, $E(Y|X=1,T=0.2)$, $Q(0.5|X=1,T=0.2)$, and $F(5|X=1, T=0.2)$. CPMs with an independence working correlation structure had minimal bias and coverage near 0.95. Estimates of $\beta_T$ from CPMs with a properly specified exchangeable working correlation structure tended to be slightly more biased ($\sim$3\%) and have lower than nominal coverage (0.91) but were much more efficient than those using independence working correlation (
empirical SE of 0.069 vs. 0.091). There was some efficiency loss fitting CPMs with an exchangeable working correlation compared to the gold standard GEE estimator that assumes the correct transformation and correlation structure (up to 31\% for $\beta_T$). Working exchangeable and independence structures yielded approximately equal precision when estimating condition quantities since estimates are based on the entire linear predictors, including the intercept function, for which working covariance weighting has a small impact on estimation efficiency. 

% alpha=0.3
When the within cluster correlation was relatively low ($\alpha=0.3$), CPMs were approximately valid with unbiased estimates of parameters and uncertainty; as expected, all relative efficiencies were close to 1.

\subsection{Equal-quantile Binning and Rounding}

\begin{table}
\caption{Simulation results for fitting CPMs with exchangeable working correlation with equal-quantile binned and rounded response data based on the primary setting. For equal-quantile binning, we show results of $M_b=50, 100$ and $200$. Results of rounding to 0 and 1 decimal place are also shown.}
    \label{tab:bin_round}

    \begin{tabular}{c c ccccc}
    \hline
    Scenario & Metric & $\beta_X$ & $\beta_T$ & ${\scriptstyle E(Y|X=1,T=0.2)}$   & ${\scriptstyle Q(0.5|X=1,T=0.2)}$ & ${\scriptstyle F(5|X=1,T=0.2)}$\\
    \hline
    \multirow{5}{*}{\parbox{6em}{Binning $M_b=50$}}  & Bias(\%) & 0.174 & 0.757 & -0.039 & -0.053 & -0.233\\
    & RMSE & 0.052 & 0.068 & 1.205 & 1.166 & 0.171\\
    & Empirical SE & 0.052 & 0.068 & 0.135 & 0.131 & 0.016 \\
    & Average SE & 0.053 & 0.067 & 0.133 & - &  0.016 \\
    & Coverage & 0.958 & 0.942 & 0.929 & 0.923 & 0.935\\
    %& & RE & 1.046 & 1.254 & 0.939 & 0.985 & 0.935\\
    \hline
 
    \multirow{5}{*}{\parbox{6em}{Binning $M_b=100$}}   & Bias(\%) & 0.187 & 1.193 & -0.316 & -0.493 & -0.174\\
    & RMSE & 0.052 & 0.069 & 1.217 & 1.181 & 0.171\\
    & Empirical SE & 0.052 & 0.068 & 0.135 & 0.130 & 0.016 \\
    & Average SE & 0.053 & 0.067 & 0.135 & - &  0.016 \\
    & Coverage & 0.957 & 0.936 & 0.945 & 0.948 & 0.953\\
    %& & RE & 1.047 & 1.261 & 0.938 & 0.975 & 0.998\\
    \hline
    
    \multirow{5}{*}{\parbox{6em}{Binning $M_b=200$}}  & Bias(\%) & 0.208 & 2.069 & -0.197 & -0.311 & -0.112\\
    & RMSE & 0.052 & 0.071 & 1.223 & 1.189 & 0.170\\
    & Empirical SE & 0.052 & 0.068 & 0.135 & 0.131 & 0.015 \\
    & Average SE & 0.053 & 0.067 & 0.136 & - &  0.016 \\
    & Coverage & 0.957 & 0.924 & 0.946 & 0.952 & 0.958\\
    %& & RE & 1.045 & 1.282 & 0.944 & 0.967 & 0.971\\
    \hline
    
    \multirow{5}{*}{\parbox{6.6em}{Rounding\\ 0 decimal place}}  & Bias(\%) & 0.196 & 0.799 & -0.015 & -7.316 & -20.965\\
    & RMSE & 0.052 & 0.070 & 1.231 & 0.940 & 0.222\\
    & Empirical SE & 0.054 & 0.069 & 0.136 & 0.155 & 0.014 \\
    & Average SE & 0.053 & 0.068 & 0.139 & - &  0.014 \\
    & Coverage & 0.959 & 0.937 & 0.952 & 0.244 & 0.004\\
    %& & RE & 1.055 & 1.319 & 0.960 & 1.375 & 0.787\\
    \hline
    
    \multirow{5}{*}{\parbox{6.6em}{Rounding\\ 1 decimal place}} & Bias(\%) & 0.210 & 3.180 & -0.123 & -0.693 & -2.147\\
    & RMSE & 0.052 & 0.076 & 1.229 & 1.175 & 0.176\\
    & Empirical SE & 0.052 & 0.070 & 0.136 & 0.130 & 0.015 \\
    & Average SE & 0.053 & 0.067 & 0.138 & - &  0.016 \\
    & Coverage & 0.957 & 0.907 & 0.953 & 0.942 & 0.943\\
    %& & RE & 1.049 & 1.325 & 0.946 & 0.971 & 0.935\\
    \hline
    \end{tabular}
\end{table}

In the primary simulation setting, when applying CPMs with exchangeable working correlation, we used equal-quantile binning with $M_b=300$. To investigate the sensitivity of results to this choice, we repeated simulations using different binning/rounding strategies. Table \ref{tab:bin_round} shows results. As $M_b$ increased, we observed fairly similar performance with slightly higher bias in coefficient estimation, especially for $\beta_T$, and slightly lower bias in conditional quantities, likely due to increasing the number of intercepts and have better estimation of the reference CDF.
% rounding
Rounding to 0 decimal place resulted in 169 categories in the response variable on average, resulting in severe information loss and poor performance for estimating $Q(0.5|X=1,T=0.2)$ and $F(5|X=1,T=0.2)$. Rounding is a sub-optimal choice for such right-skewed responses because many distinct values at the lower end of the distribution are rounded to a single value.  There were 498 ordinal levels on average if the response variable was rounded to 1 decimal place, and the performance of the estimators improved. 

\subsection{Other Simulation Results}

Results for other simulation settings are shown in Web Appendix B in Supporting Information. We give a brief summary of some other simulation results here. 
With complete data and the same time-varying covariate pattern across all subjects, CPMs with independence working correlation were as efficient as with exchangeable working correlation structure. 
When sample sizes were small, CPMs with independence working correlation exhibited good performance while CPMs with exchangeable working correlation had substantial bias; this bias decreased as the sample size increased.
When data were generated under the AR1 correlation structure, CPMs with AR1 working correlation worked well and were almost as efficient as continuous GEE methods under the correct transformation with AR1 working correlation structure.
CPMs had reasonable performance with moderate link function misspecification, i.e., when data were generated with normal residuals but fit using the logit link function.
A fully-iterated repolr procedure appeared to be slightly less biased but slightly less efficient than the one-step repolr procedure.

\section{Applications}

To illustrate the use of the proposed CPM methods, we applied them to two real data sets. The first studies CD4:CD8 ratios among people living with HIV. The second considers lung function among smokers with mild COPD.

\subsection{CD4:CD8 Ratio}

% CD4 and CD8 introduction
The CD4:CD8 ratio is the ratio of CD4 lymphocyte count (cells/mm\textsuperscript{3}) to CD8 lymphocyte count (cells/mm\textsuperscript{3}). It has been associated with immune senescence, inflammation, and comorbidities for people living with HIV \citep{castilho2016cd4}. As highlighted in the Introduction, CD4:CD8 ratio tends to be right-skewed and there is no standard transformation (shown in Web Appendix E in Supporting Information).
To study the relationship between CD4:CD8 ratio and several predictors, an observational cohort study was conducted among people living with HIV who had been on antiretroviral therapy (ART) for one year, had a suppressed viral load, and received treatment at the Vanderbilt Comprehensive Care Clinic (VCCC) between 1998 and 2012 \citep{castilho2016cd4}. In the  current analysis, we are interested in factors associated with CD4:CD8 ratio during one year of follow-up, i.e., during  the second year after starting ART. CD4:CD8 ratio was collected longitudinally during routine clinical visits. Our study included 1763 subjects with a mean of 2.9 CD4:CD8 measurements (median = 3; range = 1-7), and 3862 distinct values in the outcome. 

CPMs with independence working correlation is able to handle 3862 ordinal levels efficiently, while CPMs with exchangeable or AR1 working correlation requires binning or rounding due to computational complexities. For the latter, we divided the outcome into 1000 bins and rounded to 2 decimal places. The equal-quantile binning resulted in 979 ordinal levels due to ties on the original scale. The 2 decimal place rounding led to 234 levels. The logit link was used in all models. The time-invariant covariates considered were calendar year at baseline (one year after ART initiation), race, baseline age, sex, probable route of infection, hepatitis C virus (HCV) infection status, and hepatitis B virus (HBV) infection status. Time (in years) after baseline was the only time-varying covariate.

Odds ratio estimates and 95\% confidence intervals from the fitted CPMs are shown in Table \ref{tab:cd4cd8}. The results suggest that time, race, baseline age, and sex are associated with CD4:CD8 ratio.  For example, fixing other variables, a 10-year increase in baseline age is associated with 33\% decrease in the odds of having higher CD4:CD8 ratio based on the CPM with an independence working correlation. Results are fairly similar across all three fitted CPMs.

\begin{table}[]
    \centering
    \caption{Odds ratio estimates of higher CD4:CD8 ratios with 95\% confidence intervals from CPMs with independence working correlation and CPMs with exchangeable working correlation with binning (1000 equal-quantile bins) and rounding (2 decimal place) are shown. Variance ratios (VRs) are calculated by the variances of the log-odds ratios from CPMs with exchangeable working correlation divided by the variances of the log-odds ratios from CPMs with independence working correlation. Notably, VRs are the same up to two decimals for binning and rounding.}
    \label{tab:cd4cd8}
    \begin{tabular}{lcccc}
        \hline
        \multirow{2}{*}{Predictor} & \multirow{2}{*}{Independence} & Exchangeable &   Exchangeable & \multirow{2}{*}{VR}\\
        &  & (Binning)  & (Rounding) & \\
        \hline
        \textbf{Time} (years) &  1.22 (1.08, 1.37) & 1.23 (1.14, 1.33) &  1.23 (1.13, 1.32) & 0.43 \\
        \textbf{Enrollment Year} & 1.01 (0.98, 1.04) & 1.01 (0.99, 1.04) &  1.014 (0.99, 1.04) & 0.81 \\
        \textbf{Race} & & & &\\
        \hspace{3mm}African American  & (Reference) & & \\
        \hspace{3mm}Caucasian & 1.01 (0.83, 1.24) & 1.07 (0.89, 1.29) & 1.06 (0.88, 1.28) & 0.88\\
        \hspace{3mm}Hispanic & 0.68 (0.46, 0.99) & 0.73 (0.50, 1.06) & 0.72 (0.50, 1.05) & 0.98\\
        \hspace{3mm}Other & 0.72 (0.47, 1.12) & 0.74 (0.49, 1.12) & 0.73 (0.48, 1.11) & 0.89\\
        \textbf{Baseline Age} (10 years) & 0.67 (0.61, 0.74) & 0.68 (0.62, 0.74) & 0.68 (0.62, 0.74) & 0.88\\
        \textbf{Sex} & & & & \\ 
        \hspace{3mm}Male  & (Reference) & & & \\
        \hspace{3mm}Female & 1.72 (1.32, 2.25) & 1.80 (1.40, 2.32) &  1.80 (1.40, 2.32) & 0.90 \\
        \textbf{Route} & & & & \\ 
        \hspace{3mm}Heterosexual  & (Reference) & &  &\\
        \hspace{3mm}Injection Drug Use & 0.99 (0.68, 1.46) & 0.93 (0.64, 1.35) & 0.93 (0.64, 1.35) & 0.93 \\
        \hspace{3mm}MSM & 0.90 (0.70, 1.17) & 0.90 (0.71, 1.15) &0.90 (0.71, 1.14) & 0.89\\
        \hspace{3mm}Other/Unknown & 0.79 (0.47, 1.35) & 0.86 (0.54, 1.38) &  0.85 (0.53, 1.37) & 0.79\\
        \textbf{HCV} & 0.82 (0.60, 1.14) & 0.81 (0.60, 1.09) & 0.81 (0.60, 1.09) & 0.85\\
        \textbf{HBV} & 0.99 (0.66, 1.49) & 0.92 (0.64, 1.31) & 0.92 (0.64, 1.32) & 0.77\\
        \hline
    \end{tabular}
\end{table}

There were some differences in efficiency of estimates across different CPM estimating procedures. The variance ratios in Table \ref{tab:cd4cd8} correspond to the variances of the log-odds ratios from CPMs with exchangeable working correlation divided by the variances of the log-odds ratios from CPMs with independence working correlation. The variances for the estimated log-odds ratio for the time-varying covariate, time,  for the two exchangeable working correlation models was 0.43 times that for the independence working correlation model. We saw variance ratios ranging from 0.77 to 0.98 for time-invariant covariate parameter estimates. 

\begin{figure}
    \centering
    \includegraphics[scale=0.7]{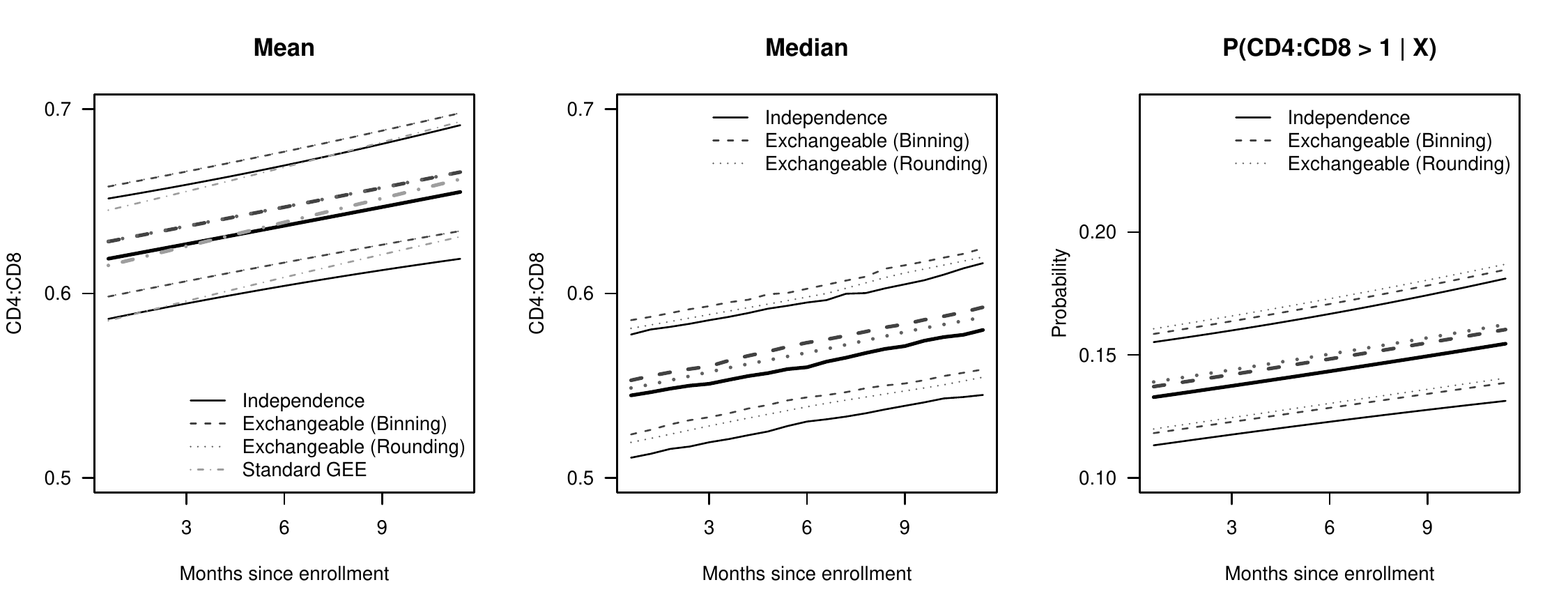}
    \caption{The estimated conditional mean CD4:CD8 ratio, median CD4:CD8 ratio, and the conditional probability that CD4:CD8 ratio  is greater than 1 as functions of months since enrollment while fixing other covariates at their medians (for continuous covariates) or modes (for categorical covariates). The estimated conditional means from the two models with exchangeable working correlation structure were almost identical.}
    \label{fig:cd4cd8_conditional}
\end{figure}

In addition to odds ratios, other quantities can be estimated from the fitted CPMs. Conditional means, and medians of CD4:CD8 and the conditional probabilities of CD4:CD8 being greater than 1 are shown as a function of time since baseline in Figure \ref{fig:cd4cd8_conditional} with other covariates fixed at their median (for continuous covariates) or mode (for categorical covariates) levels. CD4:CD8 ratio above 1 is considered normal for people without HIV \citep{petoumenos2017cd4}.  Results from the three models were generally very close. We also included the conditional mean obtained by a standard GEE model without transforming the response data for purpose of comparison; results from this analysis are also fairly similar.  

\subsection{The Lung Health Study}

The Lung Health Study was a randomized clinical trial that enrolled smokers with mild COPD from 10 centers in the United States and Canada from 1986 to 1994. The purpose of the Lung Health Study was to determine whether a smoking intervention program and the use of an inhaled bronchodilator could slow the rate of decline in lung function \citep{ anthonisen1994effects}.
For our purpose, interest was in the genetic contributions of a single nucleotide polymorphism (SNP), rs12194741, on chromosome 6 to lung function decline over 5 years \citep{hansel2013genome}. Lung function was quantified as the amount of air (in liters) one can force from the lung in the first second of exhalation (FEV1). rs12194741 was represented by a binary indicator for the presence of  at least 1 copy of the T allele. The interaction of rs12194741 and visits was used to evaluate the genetic contribution to lung function decline.  Data were collected from participants' annual visits over a 5-year follow-up period. 
In this analysis, we included participants who were continuous smokers dropping all observations after smoking stopped, and who had at least 2 observations. There were 2562 subjects included and 1694 (66\%) completed 5 visits. 
Baseline adjustment covariates included age, study site,  body mass index (BMI, weight(kg)/height($\text{m}^2$)), lifetime smoking status (in pack years), and average number of cigarettes smoked per day over the year prior to enrollment. BMI change from baseline and study visit were included as time-varying covariates. 
The distribution of the responses, FEV1, was fairly symmetric (Web Appendix F in Supporting Information). 
% baseline covariates eda (might not need so much information)

We applied both CPMs with independence and AR1 working correlation and with the logit link on the  data and compared the results. Neither binning nor rounding was applied prior to fitting the models as there were only 361 distinct values of the outcome.  Table \ref{tab:rs12194741} shows odds ratio estimates of higher FEV1 and 95\% confidence intervals obtained from the two methods. The odds ratios from the two models 
were very close. The variance ratios (VRs) shown in the last column indicate that, as expected, the log-odds ratio estimates obtained by CPMs with AR1 working correlation were more precise than those from CPMs with independence working correlation, particularly for time-varying covariates (visit and BMI change from baseline). The confidence interval for the interaction term did not cover 1, consistent with rs12194741 being associated with more rapid lung function decline at the two-sided 0.05 significance level. BMI change from baseline, baseline age, and  lifetime smoking status was negatively associated with FEV1 while baseline BMI and the average number of cigarettes smoked per day had positive associations with FEV1. For example, holding other covariates constant, a 5 kg/m$^2$ increase in BMI change from baseline was associated with a 34-35\% decrease in the odds of having a higher FEV1 value.

\begin{table}[]
    \centering
    \caption{Odds ratios estimates for higher FEV1 with 95\% confidence intervals from CPMs with independence and AR1 working correlation. The last column shows the variance ratios (VRs) calculated by the variances of the log-odds ratios from CPMs with AR1 working correlation divided by the variances the log-odds ratios from CPMs with independence working correlation.}
    \label{tab:rs12194741}
    \renewcommand{\arraystretch}{} % Default value: 1
    \begin{tabular}{lccc}
        \hline
        Predictor & Independence & AR1 & VR\\
        \hline
        \textbf{Visit} & 0.859 (0.842, 0.877) & 0.858 (0.845, 0.872) & 0.639 \\
        \textbf{rs12194741} & 1.120 (0.971, 1.291) & 1.119 (0.973, 1.287) & 0.965\\
        \textbf{Visit $\times$ rs12194741 interaction} & 0.965 (0.941, 0.989) & 0.967 (0.948, 0.986) & 0.609 \\
        \textbf{BMI Change} (per 5 kg/m\textsuperscript{2}) & 0.651 (0.529, 0.801) & 0.655 (0.561, 0.76) & 0.558 \\
        \textbf{Baseline Age} (per 10–year) & 0.342 (0.300, 0.389) & 0.341 (0.302, 0.386) & 0.895 \\
        \textbf{Baseline BMI} (per 5 kg/$\text{m}^2$) & 1.480 (1.343, 1.631) &  1.479 (1.350, 1.620) & 0.880 \\
        \textbf{Cigarettes/day} (per 10 cigs/day) & 0.976 (0.920, 1.034) & 0.975 (0.921, 1.032) & 0.956 \\
        \textbf{Pack Years} (per 20 pack year) & 1.190 (1.085, 1.304) & 1.188 (1.085, 1.301) &  0.976 \\
       \textbf{Study Site} & & &\\
       \hspace{3mm}1 & (Reference) & &\\
       \hspace{3mm}2 & 2.028 (1.429, 2.878) & 2.000 (1.449, 2.759) & 0.846\\
       \hspace{3mm}3 & 1.422 (1.001, 2.019) & 1.413 (1.021, 1.957) & 0.859 \\
       \hspace{3mm}4 & 1.811 (1.268, 2.588) & 1.807 (1.305, 2.500) & 0.829\\
       \hspace{3mm}5 & 2.671 (1.909, 3.738) & 2.636 (1.933, 3.596) & 0.853\\
       \hspace{3mm}6 & 1.950 (1.374, 2.770) & 1.919 (1.387, 2.653) & 0.856\\
       \hspace{3mm}7 & 0.908 (0.635, 1.297) & 0.907 (0.654, 1.257) & 0.837\\
       \hspace{3mm}8 &   1.724 (1.234, 2.409) & 1.703 (1.252, 2.318) & 0.849\\
       \hspace{3mm}9 & 2.016 (1.425, 2.852) & 1.987 (1.445, 2.731) & 0.840\\
       \hspace{3mm}10 & 2.307 (1.585, 3.357) & 2.292 (1.616, 3.251) &  0.868\\
    \hline
    \end{tabular}
\end{table}

Conditional quantities including means, medians, and probabilities of FEV1 being less than or equal to 2L were derived from the models and are shown in Web Appendix E in Supporting Information as a function of study visit and genotype.

\section{Discussion}

% still need to work on it!!!

We extended CPMs, a class of ordinal regression models for cross-sectional responses, to analyze clustered continuous response data. In scalar-response settings, CPMs have been used to fit different types of continuous response variables \citep{liu2017modeling, tian2020empirical}. Only rank information is used in CPMs when estimating $\bm{\beta}$, and thus fitting such ordinal regression models can avoid transformations of response variables.  To account for correlation between observations within each cluster, we estimated parameters in CPMs using GEE techniques. With the estimated parameters, we can easily obtain CDFs, expectations and quantiles conditional on covariates to help better interpret regression results.

We proposed two feasible and computationally efficient approaches for fitting CPMs depending on working correlation structures. With low within cluster correlation, CPMs with independence working correlation are able to provide unbiased estimation with proper confidence interval coverage rates and without substantial efficiency losses. With high within cluster correlation, CPMs with exchangeable/AR1 working correlation can improve efficiency. Our approaches work well under a variety of simulation settings studied for this paper. We built an \textsf{R} package, \textbf{cpmgee}, for CPMs with independence, exchangeable and AR1 working correlation. 

% limitations and future research
Our CPM methods can fit fully continuous clustered data with an independence working correlation structure, but for computational reasons  might require binning or rounding if using exchangeable or AR1 working correlation structures. For future research, we will extend CPMs to include sampling weights. With weighted CPMs, we could fit fully continuous clustered data with more complex working correlation structures by choosing different weighting matrices, and we will be able to extend the methods to address data that are missing at random, which are generally not valid with standard GEE methods.

\section*{Acknowledgements}

We would like to thank Jessica Castilho and other VCCC investigators for providing data for the HIV study. Data for The Lung Health Study were downloaded from the National Center for Biotechnology Information’s Database of Genotypes and Phenotypes (accession no. phs000335.v2.p2). This project was supported by funding from the U.S. National Institutes of Health grants R01 AI093234, R01 HL094786, R01 HL072966, P30 AI110527, and K23 AI120875. The Lung Health Study was supported by contract N01-HR-46002 from the National Heart, Lung, and Blood Institute.\vspace*{-8pt}

\bibliography{ref.bib}

\end{document}

\date{}
\maketitle

\newpage
\section{Web Appendix A}

The marginal regression model used in GEE methods for ordinal response variables is the CPM. CPMs with independence correlation and GEE methods for ordinal response variables with independence working correlation assuming observations within clusters are independent. We would like to show the estimations for $\bm{\theta}$ and $V_{\bm{\theta}}$ from CPMs with independence working correlation and GEE methods for ordinal response variables with independence working correlation are equivalent. More specifically, we first show that the score equation in CPMs is equivalent to the estimating function in GEE methods when assuming independence working correlation, then we demonstrate the equivalence of the covariance estimator.

Before directly working with the likelihood of CPMs, we first introduce some new notations. Let $O_{it,j}=I(Y_{it} = y_{(j)}) = Z_{it,j} - Z_{it,j-1}$ and $\pi_{it,j}=E(O_{it,j}|\bm{X}_{it})$. Then $\bm{O}_{it} = (O_{it,1},\dots, O_{it,j})^T$ and $\bm{O}_{it}\sim \text{Multinomial}(1, \bm{\pi}_{it})$, which belongs to the exponential family. The probability mass function (PMF) is  
\begin{align*}
    P(\bm{O}_{it}|\bm{X}_{it}) &=\left( \prod_{j=1}^{J-1} \pi_{it,j}^{O_{it,j}} \right)\left( 1 - \sum_{j=1}^{J-1} \pi_{it,j} \right)^{\left( 1 - \sum_{j=1}^{J-1} O_{it,j}\right)}\\
    &= \exp\left\{\sum_{j=1}^{J-1} O_{it,j}\log(\pi_{it,j}) + \left(1 - \sum_{j=1}^{J-1} O_{it,j}\right)\log\left( 1 - \sum_{j=1}^{J-1}\pi_{it,j} \right) \right\}\\
    &= \exp\left\{\sum_{j=1}^{J-1}O_{it,j} \log\left( \frac{\pi_{it,j}}{1 - \sum_{j=1}^{J-1}\pi_{it,j}} \right) + \log\left( 1 - \sum_{j=1}^{J-1}\pi_{it,j} \right) \right\}.
\end{align*}
The log-likelihood is 
\begin{align}
    l_O = \sum_{i=1}^N \sum_{t=1}^{T_i} \sum_{j=1}^{J-1}O_{it,j} \log\left( \frac{\pi_{it,j}}{1 - \sum_{j=1}^{J-1}\pi_{it,j}} \right) + \log\left( 1 - \sum_{j=1}^{J-1}\pi_{it,j} \right)\    \label{eq:cpm_exp_l}
\end{align}

The score equation of variables in the exponential family have a specific form \citep{mccullagh1983generalized}. Let $\bm{\pi}_i=(\bm{\pi}_{it}^T, \dots, \bm{\pi}_{iT_i}^T)^T$ and $\bm{O}_i = (\bm{O}_{i1}^T,\dots, \bm{O}_{iT_i}^T)^T$. The score equation based on (\ref{eq:cpm_exp_l}) is
\begin{align}
    U_O(\bm{\theta})=\sum_{i=1}^N \left( \frac{\partial \bm{\pi}_{i}}{\partial \bm{\theta}} \right)^T \bm{S}_{Oi}^{-1} (\bm{O}_{i} - \bm{\pi}_{i})=\bm{0},
\end{align}
where $\bm{S}_{Oi}$ is a block diagonal matrix with $\text{Cov}(\bm{O}_{it})=\text{diag}(\bm{\pi}_{it})-\bm{\pi}_{it}\bm{\pi}_{it}^T$ on the diagonal, i.e. $\bm{S}_{Oi}=\text{diag}\{\text{Cov}(\bm{O}_{i1}), \dots, \text{Cov}(\bm{O}_{iT_i})\}$.

In CPMs, we use cumulative indicators $Z_{it,j}=\sum_{k=1}^j O_{it,k}$ and cumulative probabilities $\mu_{it,j}=\sum_{k=1}^j \pi_{it,k}$. The underlying model is still multinomial and can be converted by a $(J-1)\times (J-1)$ matrix $\bm{L}=\left[\begin{matrix}1 & 0 & 0 & \dots & 0\\
1 & 1 & 0 & \dots & 0 \\
1 & 1 & 1 & \dots & 0 \\
&  \vdots & & &\\
1 & 1 & 1 & \dots & 1 \end{matrix}\right]$. The score function of CPMs can be derived by $\bm{Z}_i=\bm{L} \bm{O}_i$ and $\bm{\mu}_i = \bm{L} \bm{\pi}_i$ \citep{mccullagh1983generalized}:
\begin{align}
    U_Z(\bm{\theta})=\sum_{i=1}^N \left( \frac{\partial \bm{\mu}_{i}}{\partial \bm{\theta}} \right)^T \bm{S}_{Zi}^{-1} (\bm{Z}_{i} - \bm{\mu}_{i})=\bm{0},
    \label{eq:cpm_score}
\end{align}
where $\bm{S}_{Zi}=\text{diag}\{\text{Cov}(\bm{Z}_{i1}), \dots, \text{Cov}(\bm{Z}_{iT_i})\}$ and $\bm{S}_{Zi}=\bm{L} \bm{S}_{Oi} \bm{L}^T$.

Similarly, the information is
\begin{align}
    I_Z(\bm{\theta)}=\sum_{i=1}^N  \left( \frac{\partial \bm{\mu}_{i}}{\partial \bm{\theta}} \right)^T \bm{S}_{Zi}^{-1} \left( \frac{\partial \bm{\mu}_{i}}{\partial \bm{\theta}} \right)^T.
\end{align}
Then the robust covariance of CPMs can be estimated as
\begin{align}
\begin{split}
    \hat{V}_{\bm{\theta},\text{CPM}}=&\left\{ \sum_{i=1}^N  \left( \frac{\partial \hat{\bm{\mu}}_{i}}{\partial \bm{\theta}} \right)^T \hat{\bm{S}}_{Zi}^{-1} \left( \frac{\partial \hat{\bm{\mu}}_{i}}{\partial \bm{\theta}} \right)^T \right\}^{-1} 
    \left\{\sum_{i=1}^N \left( \frac{\partial \hat{\bm{\mu}}_{i}}{\partial \bm{\theta}} \right)^T \hat{\bm{S}}_{Zi}^{-1} (\bm{Z}_{i} - \hat{\bm{\mu}}_{i})(\bm{Z}_{i} - \hat{\bm{\mu}}_{i})^T \hat{\bm{S}}_{Zi}^{-1} \left(\frac{\partial \hat{\bm{\mu}}_{i}}{\partial \bm{\theta}}\right)  \right\} \\
    &\left\{ \sum_{i=1}^N  \left( \frac{\partial \hat{\bm{\mu}}_{i}}{\partial \bm{\theta}} \right)^T \hat{\bm{S}}_{Zi}^{-1} \left( \frac{\partial \hat{\bm{\mu}}_{i}}{\partial \bm{\theta}} \right)^T \right\}^{-1}.
    \label{eq:cov_cpm}
\end{split}
\end{align}

For GEE methods, the independence working correlation indicates that $\bm{W}_i=\bm{S}_i^{\frac{1}{2}} \bm{R}_i(\bm{\alpha})\bm{S}_i^{\frac{1}{2}} = \bm{S}_i$, where $\bm{S}_i$ is a block diagonal matrix with $\text{Cov}(\bm{Z}_{it})$ be the diagonal elements. This means $\bm{S}_i = \bm{S}_{Zi}$. The estimating equation with independence working correlation is
\begin{align}
    A_{\bm{\theta}}(\bm{\theta})= \sum_{i=1}^N \left( \frac{\partial \bm{\mu}_i}{\partial \bm{\theta}} \right)^T \bm{S}_{Zi}^{-1} (\bm{Z}_i - \bm{\mu}_i)=\bm{0}.
    \label{eq:gee_score}
\end{align}
Now (\ref{eq:cpm_score}) and (\ref{eq:gee_score}) are identical and thus solving the two equations would result in the same point estimations.

The covariance matrix in GEE methods assuming independence is estimated by
\begin{align}
\begin{split}
    \hat{V}_{\bm{\theta},\text{GEE}} = &\left\{\sum_{i=1}^N \left( \frac{\partial \hat{\bm{\mu}}_i}{\partial \bm{\theta}} \right)^T \hat{\bm{S}}_{Zi}^{-1} \left( \frac{\partial \hat{\bm{\mu}}_i}{\partial \bm{\theta}} \right) \right\}^{-1}
    \left\{\sum_{i=1}^N \left( \frac{\partial \hat{\bm{\mu}}_i}{\partial \bm{\theta}} \right)^T \hat{\bm{S}}_{Zi}^{-1} (\bm{Z}_i-\hat{\bm{\mu}}_i)(\bm{Z}_i-\hat{\bm{\mu}}_i)^T \hat{\bm{S}}_{Zi}^{-1} \left( \frac{\partial \hat{\bm{\mu}}_i}{\partial \theta} \right) \right\}\\
    & \left\{\{\sum_{i=1}^N \left( \frac{\partial \hat{\bm{\mu}}_i}{\partial \bm{\theta}} \right)^T \hat{\bm{S}}_{Zi}^{-1} \left( \frac{\partial \hat{\bm{\mu}}_i}{\partial \bm{\theta}} \right) \right\}^{-1}.
    \label{eq:cov_gee}
\end{split}
\end{align}
(\ref{eq:cov_cpm}) and (\ref{eq:cov_gee}) are also identical. Therefore, we have shown that CPMs with independence working correlation is equivalent to GEE methods for ordinal response variables with independence working correlation.

\section{Web Appendix B}

\citet{li2022asymptotics} has shown consistency and asymptotic normality for NPMLEs in CPMs in cross-sectional settings under mild conditions including boundedness of the response variable. The proof for CPMs with independence working correlation on response data censored at a lower bound $L$ and an upper bound $U$, where the bounds satisfy $\Pr(L<Y<U)>0$, $\Pr(Y\le L)>0$, and $\Pr(Y\ge U)>0$ is very similar as the proof in \citet{li2022asymptotics} with minor modifications to address for correlated responses and sandwich estimator for covariance. We use the same notation in \citet{li2022asymptotics} ($\bm{\gamma}$, $\bm{X}$, and $G^{-1}$ in this paper are equivalent to $A$, $Z$, and $G$ in Li's paper respectively).

Suppose there are $n$ subjects, and subject $i$ has $T_i$ observations ($i=1,\dots,n$).  Let $Y_{it}$ be the outcome for subject $i$ at time $t$. Let $J$ be the number of distinct values in the observed outcomes $\{Y_{it}\}$, and let $y_{(j)}$ be the $j$-th smallest value among the $J$ distinct values.  Let $Z_{it}$ be the vector of covariates for subject $i$ at time $t$.  The linear transformation model for such clustered data is
\begin{equation*}
  A(Y_{it})=\beta^T Z_{it} + \epsilon_{it}, \quad \epsilon_{it}\sim G,
  \label{eq:linear_transformation_model}
\end{equation*}
Here, $\epsilon_{it}$ and $\epsilon_{i't'}$ are independent when $i\neq i'$, but they may not be independent when $i=i'$.  This model is equivalent to the CPM, \begin{equation*}
  G^{-1}\left\{\Pr(Y_{it}\le y_{(j)}|Z_{it})\right\} = A(y_{(j)})-\beta^T Z_{it}.
\end{equation*}

We now give sketch proofs of the asymptotic properties for CPMs on data censored at $L$ and $U$ with independence working correlation.  The proofs are very similar to those in Sections A.1 and A.2, with minor modifications to address correlated responses and the sandwich estimator for covariance. With the independence working correlation, the pseudo log-likelihood for the censored clustered data is
\begin{align*}
  l_n(\beta,A)=
  &\frac{1}{n}\sum_{i=1}^n \frac{1}{T_i}\sum_{t=1}^{T_i} \{I(Y_{it}\le L)\log G(A(L)-\beta^TZ_{it}) \\
  & \qquad\qquad\qquad +I(Y_{it}\ge U)\log(1-G(A(U-)-\beta^TZ_{it})) \\
  & \qquad\qquad\qquad +I(L<Y_{it}<U)\log(G(A(Y_{it})-\beta^T Z_{it}) - G(A(Y_{it}-)-\beta^TZ_{it}))\}.
\end{align*}

For the proof of consistency for $(\widehat\beta, \widehat A)$, the boundedness of $\widehat{A}(y)$ and $n\widehat A\{Y_i\}$ still holds following the same proof in Section A.1.  The marginal likelihood under independence working correlation is still a valid likelihood for which the Kullback--Leibler property holds.  Thus the consistency of $(\widehat\beta, \widehat A)$ holds following the same arguments as in Section A.1.

For the proof of the asymptotic distribution of $(\widehat\beta, \widehat A)$, note that equation (A.8) in Section A.2 still holds when the operators $\bm{S}_{11}$, $\bm{S}_{12}$, $\bm{S}_{21}$, $\bm{S}_{22}$ on the left-hand side are defined as the second order differentiation operators based on the pseudo log-likelihood above and $\bm{S}(Y,Z)[\nu, h]$ on the right-hand side is defined as the first order differentiation operator of the pseudo log-likelihood. As the operator $(\bm{S}_{11}^T\nu+\bm{S}_{12}[h], \bm{S}_{21}^T\nu+\bm{S}_{22}[h])$ is defined for the marginal likelihood under independence working correlation, its invertibility can be shown in a manner similarly to the one treating all data as
independent.  Thus (A.10) holds:
$$\sqrt n {\nu^*}^T(\widehat\beta-\beta_0)
+\sqrt n\int h^*(y) d(\widehat A-A_0)(y) = \sqrt n ({\bf P}_n-{\bf
P})\bm{S}(Y,Z)[\nu^-, h^-]+o_p(1).\eqno(A.10)$$ Since $v^-$ and $h^-$ are the inverse of the information operator and $\bm{S}(Y,Z)$ is the the first derivative, the asymptotic variance takes the sandwiched form, $A^{-1}E[SS'] A^{-1}$, where $A$ is the information matrix (with $\beta$ and $A$ as parameters) and $S$ is the first derivative of the pseudo log-likelihood.  Since we estimate $E[SS']$ by its empirical moment and $S$ is differentiable with respect to $\beta$ and $A$ (so is Glivenko--Cantelli), its estimator is also consistent.  Thus, the sandwiched variance is consistent.

\section{Web Appendix C}

\subsection{Complete Data}

\begin{table}
\caption{Simulation results for the complete data scenarios based on the primary setting.}
    \label{tab:complete}

    \begin{tabular}{c c c ccccc}
    \hline
    $\alpha$ & Method & Metric & $\beta_X$ & $\beta_T$ & ${\scriptstyle E(Y|X=1,T=0.2)}$   & ${\scriptstyle Q(0.5|X=1,T=0.2)}$ & ${\scriptstyle F(5|X=1,T=0.2)}$\\
    \hline
    \multirow{18}{*}{$0.7$} & & Bias(\%) & -0.006 & 0.063 & - & - & -\\
    &  & RMSE & 0.050 & 0.038 & - & - & -\\
    & GEE & Empirical SE & 0.050 & 0.038 &  - & - & -\\
    & (ex) & Average SE & 0.049 & 0.038 &  - & - & -\\
    &  & Coverage & 0.950 & 0.945 & - & - & -\\
    & & RE & \text{reference} & \text{reference} & - & - & -\\
    \cline{2-8}
    & & Bias(\%) & 0.157 & 0.300 & 0 & -0.038 & -0.122\\
    && RMSE & 0.051 & 0.043 & 1.233 & 1.200 & 0.170\\
    & CPM & Empirical SE & 0.051 & 0.043 & 0.132 & 0.127 & 0.015\\
    & (ind) & Average SE & 0.051 & 0.042 & 0.134 & - & 0.015\\
    & & Coverage & 0.945 & 0.950 & 0.953 & 0.956 & 0.950\\
    & & RE & 1.041 & 1.251 & - & - & -\\
    \cline{2-8}
    & & Bias(\%) & 0.277 & 2.955 & -0.327 & -0.394 & -0.303\\
    && RMSE & 0.051 & 0.053 & 1.218 & 1.185 & 0.169\\
    & CPM & Empirical SE & 0.051 & 0.044 & 0.131 & 0.127 & 0.015 \\
    & (ex) & Average SE & 0.051 & 0.042 & 0.132 & - & 0.015\\
    & & Coverage & 0.949 & 0.888 & 0.952 & 0.953 & 0.951\\
    & & RE & 1.051 & 1.317 & - & - & -\\
    \hline
    \multirow{18}{*}{$0.3$} &  & Bias(\%) & -0.051 & 0.073 & - & - & -\\
    & & RMSE & 0.037 & 0.058 & - & - & -\\
    & GEE & Empirical SE & 0.037 & 0.058 & - & - & -\\
    & (ex) & Average SE & 0.037 & 0.057 & - & - & -\\
    &  & Coverage & 0.942 & 0.949 & - & - & -\\
    & & RE & \text{reference} & \text{reference} & - & - & - \\
    \cline{2-8}
    &  & Bias(\%) & 0.072 & 0.207 & -0.006 & -0.031 & -0.058\\
    & & RMSE & 0.038 & 0.056 & 1.235 & 1.204 & 0.171\\
    & CPM & Empirical SE & 0.038 & 0.056 & 0.101 & 0.099 & 0.012\\
    & (ind) & Average SE & 0.037 & 0.057 & 0.103 & - & 0.012\\
    & & Coverage & 0.942 & 0.947 & 0.959 & 0.955 & 0.956\\
    & & RE & 1.035 & 0.953 & - & - & -\\
    \cline{2-8}
    & & Bias(\%) & 0.183 & 2.899 & -0.338 & -0.387 & -0.402\\
    & & RMSE & 0.038 & 0.065 & 1.219 & 1.188 & 0.170\\
    & CPM & Empirical SE & 0.038 & 0.058 & 0.101 & 0.100 & 0.012 \\
    & (ex) & Average SE & 0.038 & 0.057 & 0.103 & - & 0.012\\
    &  & Coverage & 0.943 & 0.912 & 0.953 & 0.945 & 0.948\\
    & & RE & 1.075 & 1.002 & - & - & -\\
    \hline
    
\end{tabular}
\end{table}

In an ideal situation, no value is missing. With complete data and the same time-varying covariate pattern across all subjects, each observation contributes approximately equally to the estimating equation, so the independence working correlation structure is as efficient as a more complex working correlation structure \citet{lipsitz1994analysis}.

We evaluated the performances of the two CPM methods with different association parameter $\alpha$ when we have complete data. The results are in Table \ref{tab:complete}. We do not expect and did not observe efficiency gain by using exchangeable working correlation with complete data. The CPM methods with independence working correlation had slightly better performance under this circumstance for its lower bias, more proper coverage rates, and similar RMSE. The CPM method was almost as efficient as the GEE method for continuous response variables with the correct transformation when the within cluster correlation is small. 

\subsection{Time Effects}

\begin{table}
\caption{Simulation results for different time effects ($\beta_T=0, 0.5$ and $2$).}
    \label{tab:time_effects}

    \begin{longtable}{c c c ccccc}
    \hline
    $\beta_T$ & Method & Metric & $\beta_X$ & $\beta_T$ & ${\scriptstyle E(Y|X=1,T=0.2)}$   & ${\scriptstyle Q(0.5|X=1,T=0.2)}$ & ${\scriptstyle F(5|X=1,T=0.2)}$\\
    \hline
    \multirow{16}{*}{$0$} &  & Bias(\%) & -0.01 & $\infty$ & - & - & -\\
    & GEE & RMSE & 0.710 & 0.710 & - & - & -\\
    & (ex) & Coverage & 0.953 & 0.944 & - & - & -\\
    & & RE & \text{reference} & \text{reference} & - & - & -\\
    \cline{2-8}
    & & Bias(\%) & 0.125 &  $\infty$ & -0.009 & -0.079 & -0.103\\
    & & RMSE & 0.712 & 0.711 & 1.184 & 1.150 & 0.173\\
    & CPM & Empirical SE & 0.054 & 0.089 & 0.137 & 0.128 & 0.016 \\
    & (ind) & Average SE & 0.055 & 0.086 & 0.139 & - & 0.017\\
    &  & Coverage & 0.958 & 0.947 & 0.954 & 0.959 & 0.956\\
    & & RE & 1.130 & 2.166 & - & - & - \\
    \cline{2-8}
    & & Bias(\%) & 0.157 & $\infty$  & -0.121 & -0.210 & -0.144\\
    & & RMSE & 0.712 & 0.709 & 1.180 & 1.145 & 0.174\\
     & CPM & Empirical SE & 0.052 & 0.066 & 0.133 & 0.126 & 0.016 \\
    & (ex) & Average SE & 0.053 & 0.062 & 0.134 & - & 0.017\\
    &  & Coverage & 0.959 & 0.940 & 0.945 & 0.956 & 0.956\\
    & & RE & 1.046 & 1.180 & - & - & -\\
    \hline
    
   \multirow{16}{*}{$0.5$} & & Bias(\%) & -0.010 & 0.174 & - & - & -\\
    & GEE & RMSE & 0.358 & 0.359 & - & - & -\\
    & (ex) & Coverage & 0.953 & 0.944 & - & - & -\\
    & & RE & \text{reference} & \text{reference} & - & - & -\\
    \cline{2-8}
    & & Bias(\%) & 0.126 & 0.275 & -0.009 & -0.074 & -0.089\\
    &  & RMSE & 0.361 & 0.363 & 1.208 & 1.174 & 0.172\\
    & CPM & Empirical SE & 0.054 & 0.089 & 0.138 & 0.130 & 0.016\\
    & (ind) & Average SE & 0.055 & 0.086 & 0.140 & - & 0.017\\
    &  & Coverage & 0.957 & 0.944 & 0.955 & 0.957 & 0.957\\
    & & RE & 1.130 & 2.189 & - & - & -\\
    \cline{2-8}
    & & Bias(\%) & 0.180 & 3.111 & -0.146 & -0.236 & -0.047\\
    &  & RMSE & 0.369 & 0.349 & 1.202 & 1.168 & 0.172\\
    & CPM & Empirical SE & 0.052 & 0.066 & 0.134 & 0.128 & 0.016 \\
    & (ex) & Average SE & 0.053 & 0.063 & 0.135 & - & 0.016\\
    & & Coverage & 0.95 & 0.933 & 0.941 & 0.954 & 0.958\\
    & & RE & 1.046 & 1.204 & - & - & -\\
    \hline
    
    \multirow{16}{*}{$2$} & & Bias(\%) & -0.020 & 0.047 & - & - & -\\
    & GEE & RMSE & 0.710 & 0.711 & - & - & -\\
    & (ex) & Coverage & 0.953 & 0.944 & - & - & -\\
    & & RE & \text{reference} & \text{reference} & - & - & -\\
    \cline{2-8}
    & & Bias(\%) & 0.132 & 0.279 & -0.013 & -0.086 & -0.188\\
    &  & RMSE & 0.708 & 0.727 & 1.285 & 1.255 & 0.165\\
    & CPM & Empirical SE & 0.054 & 0.098 & 0.141 & 0.137 & 0.015\\
    & (ind) & Average SE & 0.055 & 0.096 & 0.145 & - & 0.015\\
    &  & Coverage & 0.960 & 0.945 & 0.959 & 0.956 & 0.955\\
    & & RE & 1.130 & 2.637 & - & - & - \\
    \cline{2-8}
    & & Bias(\%) & 0.411 & 2.766 & -0.266 & -0.358 & -0.147\\
    & & RMSE & 0.706 & 0.751 & 1.272 & 1.242 & 0.165\\
    & CPM & Empirical SE & 0.052 & 0.079 & 0.138 & 0.136 & 0.015 \\
    & (ex) & Average SE & 0.054 & 0.078 & 0.140 & - & 0.015\\
    &  & Coverage & 0.957 & 0.888 & 0.951 & 0.947 & 0.949\\
    & & RE & 1.061 & 1.728 & - & - & -\\
    \hline
\end{longtable}
\end{table}

We varied the coefficient for time, $\beta_T$, from 0 to 2 to investigate the performance under scenarios with different time effects. Results are in Table \ref{tab:time_effects}. When $\beta_T=0$, the percent bias for both methods was $\infty$ because the true value (in the denominator) is 0, and the bias for the all methods was small (0.0009, 0.0001, and -0.0009). As the time effects increase, CPM methods were less efficient than the standard GEE method with the correct transformation, but they still had small bias and good coverage rates.

\subsection{Sample Size and Cluster Size}

\begin{table}[]
    \centering
    \caption{Simulation results for different sample sizes ($N=100, 200$ and $500$) keeping the maximum cluster size at $6$.}
    \begin{tabular}{c c c ccccc}
    \hline
          $N$ & Method & Metric & $\beta_X$ & $\beta_T$ & ${\scriptstyle E(Y|X=1,T=0.2)}$   & ${\scriptstyle Q(0.5|X=1,T=0.2)}$ & ${\scriptstyle F(5|X=1,T=0.2)}$\\
    \hline
    \multirow{16}{*}{$100$} &  & Bias(\%) & 0.551 & -0.313 & - & - & -\\
    & GEE & RMSE & 0.166 & 0.184 & - & - & -\\
    & (ex) & Coverage & 0.917 & 0.952 & - & - & -\\
    & & RE & \text{reference} & \text{reference} & - & - & -\\
    \cline{2-8}
    & & Bias(\%) & 2.114 &  1.504 & 0.289 & 0.469 & -0.757\\
    & & RMSE & 0.182 & 0.288 & 1.325 & 1.299 & 0.180\\
    &  CPM & Empirical SE & 0.181 & 0.287 &  0.464 & 0.438 & 0.052\\
    & (ind) & Average SE & 0.175 & 0.283 & 0.445 & - & 0.051\\
    & & Coverage & 0.940 & 0.945 & 0.939 & 0.939 & 0.947\\
    & & RE & 1.196 & 2.433 & - & - & - \\
    \cline{2-8}
    & & Bias(\%) & 3.122 & 40.440  & -1.114 & -0.873 & 0.985\\
    & & RMSE & 0.181 & 0.516 & 1.260 & 1.233 & 0.175\\
    & CPM & Empirical SE & 0.178 & 0.320 & 0.448 & 0.427 & 0.052 \\
    & (ex) & Average SE & 0.168 & 0.222 & 0.417 & - & 0.050\\
    & & Coverage & 0.920 & 0.556 & 0.920 & 0.942 & 0.946\\
    & & RE & 1.159 & 3.016 & - & - & -\\
    \hline
    
     \multirow{16}{*}{$200$}  &  & Bias(\%) & 0.497 & 0.099 & - & - & -\\
    & GEE & RMSE & 0.112 & 0.131 & - & - & -\\
    & (ex) & Coverage & 0.938 & 0.948 & - & - & -\\
    & & RE & \text{reference} & \text{reference} & - & - & -\\
    \cline{2-8}
    & & Bias(\%) & 1.090 &  0.873 & 0.085 & 0.164 & -0.505\\
    &  & RMSE & 0.126 & 0.195 & 1.266 & 1.240 & 0.174\\
    & CPM & Empirical SE & 0.126 & 0.195 & 0.318 & 0.307 & 0.037 \\
    & (ind) & Average SE &  0.125 & 0.198 & 0.315 & - & 0.036\\
    & & Coverage & 0.940 & 0.952 & 0.939 & 0.946 & 0.934\\
    & & RE & 1.263 & 2.200 & - & - & -\\
    \cline{2-8}
    & & Bias(\%) & 1.595 & 16.625 & -0.505 & -0.418 & 0.179\\
    & & RMSE & 0.122 & 0.242 & 1.251 & 1.221 & 0.174\\
    & CPM & Empirical SE & 0.121 & 0.176 & 0.308 & 0.300 & 0.036\\
    & (ex) & Average SE & 0.119 & 0.151 & 0.301 & - & 0.036\\
    &  & Coverage & 0.946 & 0.790 & 0.933 & 0.939 & 0.939\\
    & & RE & 1.166 & 1.797 & - & - & -\\
    \hline
    
     \multirow{16}{*}{$500$}&  & Bias(\%) & -0.259 & 0.034 & - & - & -\\
    & GEE & RMSE & 0.074 & 0.084 & - & - & -\\
    & (ex) & Coverage & 0.940 & 0.949 & - & - & -\\
    & & RE & \text{reference} & \text{reference} & - & - & - \\
    \cline{2-8}
    & & Bias(\%) & 0.074 &  0.444 & -0.049 & -0.068 & -0.035\\
    & & RMSE & 0.082 & 0.122 & 1.236 & 1.210 & 0.171\\
    & CPM & Empirical SE & 0.082 & 0.122 & 0.203 & 0.199 & 0.024\\
    & (ind) & Average SE & 0.078 & 0.125 & 0.200 & - & 0.023\\
    & & Coverage & 0.941 & 0.955 & 0.942 & 0.941 & 0.944\\
    & & RE & 1.211 & 2.137 & - & - & -\\
    \cline{2-8}
    & & Bias(\%) & 0.192 & 5.862 & -0.318 & -0.360 & 0.262\\
    & & RMSE & 0.079 & 0.114 & 1.223 & 1.196 & 0.170\\
    & CPM & Empirical SE & 0.079 & 0.098 & 0.199 & 0.197 & 0.024\\
    & (ex) & Average SE & 0.075 & 0.094 & 0.193 & - & 0.023\\
    & & Coverage & 0.940 & 0.911 & 0.927 & 0.934 & 0.931\\
    & & RE & 1.134 & 1.370 & - & - & -\\
    \hline
    \end{tabular}
    \label{tab:sample_size}
\end{table}

\begin{table}[]
    \centering
    \caption{Simulation results for different cluster sizes ($M=3$ and $12$) keeping the sample size at $1000$.}
    \begin{tabular}{c c c ccccc}
    \hline
          $M$ & Method & Metric & $\beta_X$ & $\beta_T$ & ${\scriptstyle E(Y|X=1,T=0.2)}$   & ${\scriptstyle Q(0.5|X=1,T=0.2)}$ & ${\scriptstyle F(5|X=1,T=0.2)}$\\
    \hline
    \multirow{16}{*}{$3$} & & Bias(\%) & 0.011 & -0.013 & - & - & -\\
    & GEE & RMSE & 0.052 & 0.142 & - & - & -\\
    & (ex) & Coverage & 0.950 & 0.943 & - & - & -\\
    & & RE & \text{reference} & \text{reference} & - & - & -\\
    \cline{2-8}
    &  & Bias(\%) & 0.281 &  0.157 & -0.006 & -0.049 & -0.267\\
    & & RMSE & 0.054 & 0.167 & 1.233 & 1.201 & 0.171\\
    & CPM & Empirical SE & 0.054 & 0.167 & 0.147 & 0.140 & 0.016 \\
    & (ind) & Average SE & 0.055 & 0.165 & 0.147 & - & 0.017\\
    &  & Coverage & 0.956 & 0.950 & 0.944 & 0.957 & 0.957\\
    & & RE & 1.067 & 1.387 & - & - & - \\
    \cline{2-8}
    & & Bias(\%) & 0.283 & 2.863 & -0.083 & -0.144 & -0.364\\
    & & RMSE & 0.053 & 0.154 & 1.229 & 1.196 & 0.171\\
    & CPM & 0.053 & 0.152 & 0.144 & 0.138 & 0.016 \\
    & (ex) & 0.055 & 0.145 & 0.143 & - &  0.017 \\
    & & Coverage & 0.955 & 0.938 & 0.945 & 0.956 & 0.957\\
    & & RE & 1.050 & 1.137 & - & - & -\\
    \hline
    
    \multirow{16}{*}{$12$} & & Bias(\%) & 0.041 & -0.083 & - & - & -\\
    & GEE & RMSE & 0.050 & 0.023 & - & - & -\\
    & (ex) & Coverage & 0.950 & 0.954 & - & - & -\\
    & & RE & \text{reference} & \text{reference} & - & - & -\\
    \cline{2-8}
    & & Bias(\%) & 0.221 &  0.029 & 0.026 & -0.022 & -0.212\\
    & & RMSE & 0.056 & 0.046 & 1.235 & 1.203 & 0.171\\
    & CPM & Empirical SE & 0.056 & 0.046 & 0.138 & 0.133 & 0.016 \\
    & (ind) & Average SE & 0.056 & 0.046 & 0.139 & - & 0.016 \\
    &  & Coverage & 0.949 & 0.954 & 0.955 & 0.945 & 0.950\\
    & & RE & 1.235 & 4.075 & - & - & -\\
    \cline{2-8}
    & & Bias(\%) & 0.461 & 2.455 & -0.365 & -0.401 & -0.144\\
    & & RMSE & 0.053 & 0.043 & 1.216 & 1.186 & 0.170\\
    & CPM & Empirical SE & 0.053 & 0.035 & 0.133 & 0.131 & 0.016\\
    & (ex) & Average SE & 0.053 & 0.034 & 0.133 &  - & 0.016\\
    &  & Coverage & 0.947 & 0.886 & 0.940 & 0.945 & 0.954\\
    & & RE & 1.115 & 2.310 & - & - & -\\
    \hline
    \end{tabular}
    \label{tab:cluster_size}
\end{table}

%CPMs work well with small sample sizes while GEE methods usually require larger sample sizes for good performance.

We conducted additional simulations varying the number of clusters, $N$, from 100 to 500. The cluster size is of interest as well. Let $M=\max\{T_i\}$ be the largest cluster size. Performances of the two methods were evaluated with smaller ($M=3$) and larger ($M=12$) cluster sizes while other settings were the same as the primary settings ($N=1000, M=6$). Results are shown in Table \ref{tab:sample_size} and Table \ref{tab:cluster_size}. When $N=100$, CPMs with independence working correlation had good performance while CPMs with exchangeable working correlation had substantial bias. The bias decreased and efficiency gains increased as the sample size increased. With large $N$, performance of CPMs was good regardless of cluster size. However, the RE of standard GEE over CPMs seemed to be greater as the number of clusters increased.

% Need to work on this paragraph!!!!

\subsection{First-order Autoregressive (AR1) Correlation Structure}

\begin{table}
\centering
\caption{Simulation results of fitting standard GEE and CPMs with the AR1 correlation structure. We compare the results to CPMs with independence and exchangeable working correlation structures. }
    \label{tab:ar1}

    \begin{tabular}{c c ccccc}
    \hline
    Method & Metric & $\beta_X$ & $\beta_T$ & ${\scriptstyle E(Y|X=1,T=0.2)}$   & ${\scriptstyle Q(0.5|X=1,T=0.2)}$ & ${\scriptstyle F(5|X=1,T=0.2)}$\\
    \hline
    & Bias(\%) & -0.065 & 0.144 & - & - & -\\
    & RMSE & 0.046 & 0.099 & - & - & -\\
    GEE & Empirical SE & 0.046 & 0.101 & - & - & -\\
    (AR1) & Average SE & 0.046 & 0.099 & - & - & -\\
     & Coverage & 0.954 & 0.943 & - & - & -\\
    & RE & \text{reference} & \text{reference} & - & - & -\\
    \hline
    & Bias(\%) & 0.087 & 0.277 & 0.000 & -0.062 & -0.111\\
     & RMSE & 0.048 & 0.109 & 1.231 & 1.198 & 0.170\\
    CPM & Empirical SE & 0.048 & 0.109 & 0.128 & 0.125 & 0.014 \\
    (ind) & Average SE & 0.050 & 0.109 & 0.131 & - & 0.015\\
     & Coverage & 0.960 & 0.946 & 0.958 & 0.958 & 0.950\\
    & RE & 1.115 & 1.169 &  - & - & -\\
    \hline
    & Bias(\%) & 0.089 & 0.530 & -0.107 & -0.190 & -0.114\\
   & RMSE & 0.048 & 0.103 & 1.226 & 1.193 & 0.170\\
   CPM & Empirical SE & 0.048 & 0.103 & 0.125 & 0.123 & 0.014 \\
   (AR1) & Average SE & 0.049 & 0.103 & 0.128 & - & 0.015\\
    & Coverage & 0.958 & 0.946 & 0.950 & 0.951 & 0.950\\
    & RE & 1.079 & 1.051 &  - & - & -\\
    \hline
    & Bias(\%) & 0.203 & 3.007 & -0.176 & -0.256 & -0.033\\
   & RMSE & 0.048 & 0.107 & 1.223 & 1.190 & 0.170\\
   CPM & Empirical SE & 0.048 & 0.103 & 0.126 & 0.124 & 0.015\\
   (ex) & Average SE & 0.049 & 0.103 & 0.129 & - & 0.015\\
    & Coverage & 0.961 & 0.946 & 0.950 & 0.947 & 0.958\\
    & RE & 1.004 & 1.040 &  - & - & -\\
    \hline
    
\end{tabular}
\end{table}

We generated residuals with AR1 correlation structure with $\alpha=0.7$, and fit both AR1 and exchangeable working correlation structures keeping other settings the same as the primary setting. The results are in Table \ref{tab:ar1}. CPM methods were almost as efficient as continuous GEE methods, especially with the correct AR1 working correlation structure. If fitting exchangeable working correlation, CPMs method still had small bias and correct coverage rates.

\subsection{Link Function Misspecification}
We look into the performance of our approaches with link function misspecification. The residuals were generated with standard normal distributions and we still fit models with the logit link. Results are shown in Table \ref{tab:mis_link}. Regression parameters were transformed to the same scale. CPMs methods are generally robust to moderate link function misspecification \citet{liu2017modeling, tian2020empirical}. The bias of regression parameters is larger than that in correctly specified models. Mean and median estimation are still good. The results for CDF is less satisfying under link function misspecification, with larger bias suboptimal coverage of 95\% confidence intervals.

\begin{table}
\centering
\caption{Simulation results for the link function misspecification. Data were generated based on the probit link while fitting models with the logit link. Regression parameters were transformed to comparable scales.}
    \label{tab:mis_link}

    \begin{tabular}{c c ccccc}
    \hline
     Method & Metric & $\beta_X$ & $\beta_T$ & ${\scriptstyle E(Y|X=1,T=0.2)}$   & ${\scriptstyle Q(0.5|X=1,T=0.2)}$ & ${\scriptstyle F(5|X=1,T=0.2)}$\\
    \hline
    % & Bias(\%) & -0.009 & 0.057 & - & - & -\\
    % & RMSE & 0.028 & 0.033 & - & - & -\\
    % GEE & Empirical SE & 0.028 & 0.033 &- & - & -\\
    % (ex) & Average SE & 0.028 & 0.032 & - & - & -\\
    %  & Coverage & 0.952 & 0.944 & - & - & -\\
    % & RE & \text{reference} & \text{reference} & - & - & -\\
    % \hline
     & Bias(\%) & -3.983	 & -3.879 & 0.272 & 0.503 & -6.193\\
    & RMSE & 0.052 & 0.066 & 1.220 & 1.223 & 0.271\\
    CPM & Empirical SE & 0.033 & 0.054 & 0.078 & 0.084 & 0.013 \\
    (ind) & Average SE & 0.034 & 0.052 & 0.080 & - & 0.014\\
     & Coverage & 0.793 & 0.876 & 0.956 & 0.953 & 0.842\\
    & RE & 1.414 & 2.654 & - & - & -\\
    \hline
    & Bias(\%) & -4.602 & -2.726 & 0.237 & 0.342 & -5.338\\
    & RMSE & 0.056 & 0.050 & 1.219 & 1.216 & 0.270\\
    CPM & Empirical SE & 0.032 & 0.042 & 0.077 & 0.084 & 0.013\\
    (ex)& Average SE & 0.033 & 0.040 & 0.076 & - & 0.013\\
     & Coverage & 0.722 & 0.890 & 0.952 & 0.954 & 0.864\\
    & RE & 1.341 & 1.597 & - & - & -\\
    \hline
\end{tabular}
\end{table}

\subsection{Fully-iterated repolr}

We run simulations with fully-iterated repolr methods to compare the performance of fully-iterated repolr, which can be very time-consuming, and our more efficient one-step repolr method. Fully-iterated repolr methods provided less biased results but less efficient regression parameter estimates compared to the one-step repolr method, particularly for the time-varying covariate.

\begin{table}
\centering
\caption{Simulation results for the fully-iterated repolr method with exchangeable working correlation structure based on the primary setting. We computed the efficiency relative to standard GEE methods with exchangeable working correlation structure.}
    \label{tab:fully_iterated}

    \begin{tabular}{c c ccccc}
    \hline
     Metric & $\beta_X$ & $\beta_T$ & ${\scriptstyle E(Y|X=1,T=0.2)}$   & ${\scriptstyle Q(0.5|X=1,T=0.2)}$ & ${\scriptstyle F(5|X=1,T=0.2)}$\\
    \hline
    Bias(\%) & 0.138 & 0.302 & -0.101 & -0.196 & -0.175\\
    RMSE & 0.053 & 0.083 & 1.228 & 1.194 & 0.171\\
    Empirical SE & 0.053 & 0.081 & 0.138 & 0.131 & 0.015\\
    Average SE & 0.054 & 0.083 & 0.139 & - & 0.016\\
    Coverage & 0.961 & 0.943 & 0.953 & 0.956 & 0.959\\
    RE & 1.097 & 1.901 & - & - & -\\
    \hline
\end{tabular}
\end{table}

\section{Web Appendix D}

\begin{figure}
    \centering
    \includegraphics[scale=0.72]{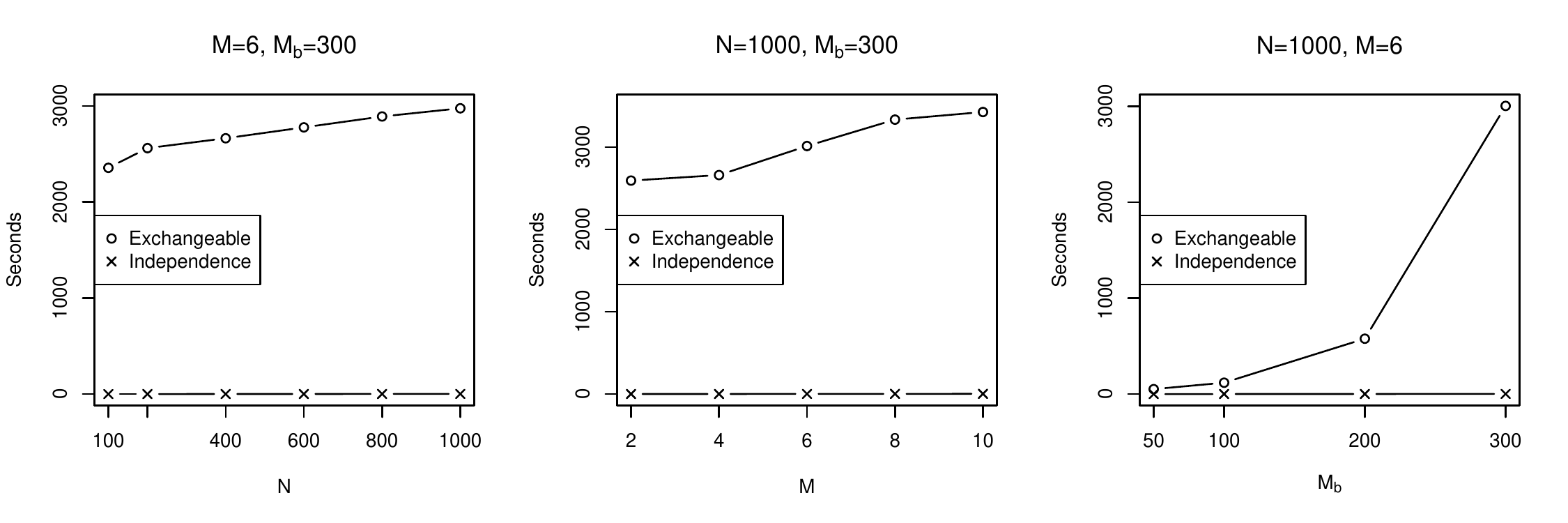}
    \caption{Computation time of CPMs with independence and exchangeable working correlation under the primary setting. The first plot shows the results when varying the sample size from 100 to 1000. The second plot show the computation time if increasing the cluster size from 2 to 10. In the third plot, we show the computation time varying the binning size from 50 to 300.}
    \label{fig:computation_time}
\end{figure}

We show the computation time of CPMs with independence and exchangeable working correlation structure under the primary simulation setting varying one of the sample size, cluster size, and binning size. Results are the average time based on 100 replications. As shown in Figure \ref{fig:computation_time}, CPMs with independence working correlation is extremely computationally  efficient. With exchangeable working correlation structure, increasing the number of distinct values in the response variable can greatly increase computation time.  %This simulation and all other simulations were performed on ...

\begin{figure}
    \centering
    \includegraphics[scale=0.7]{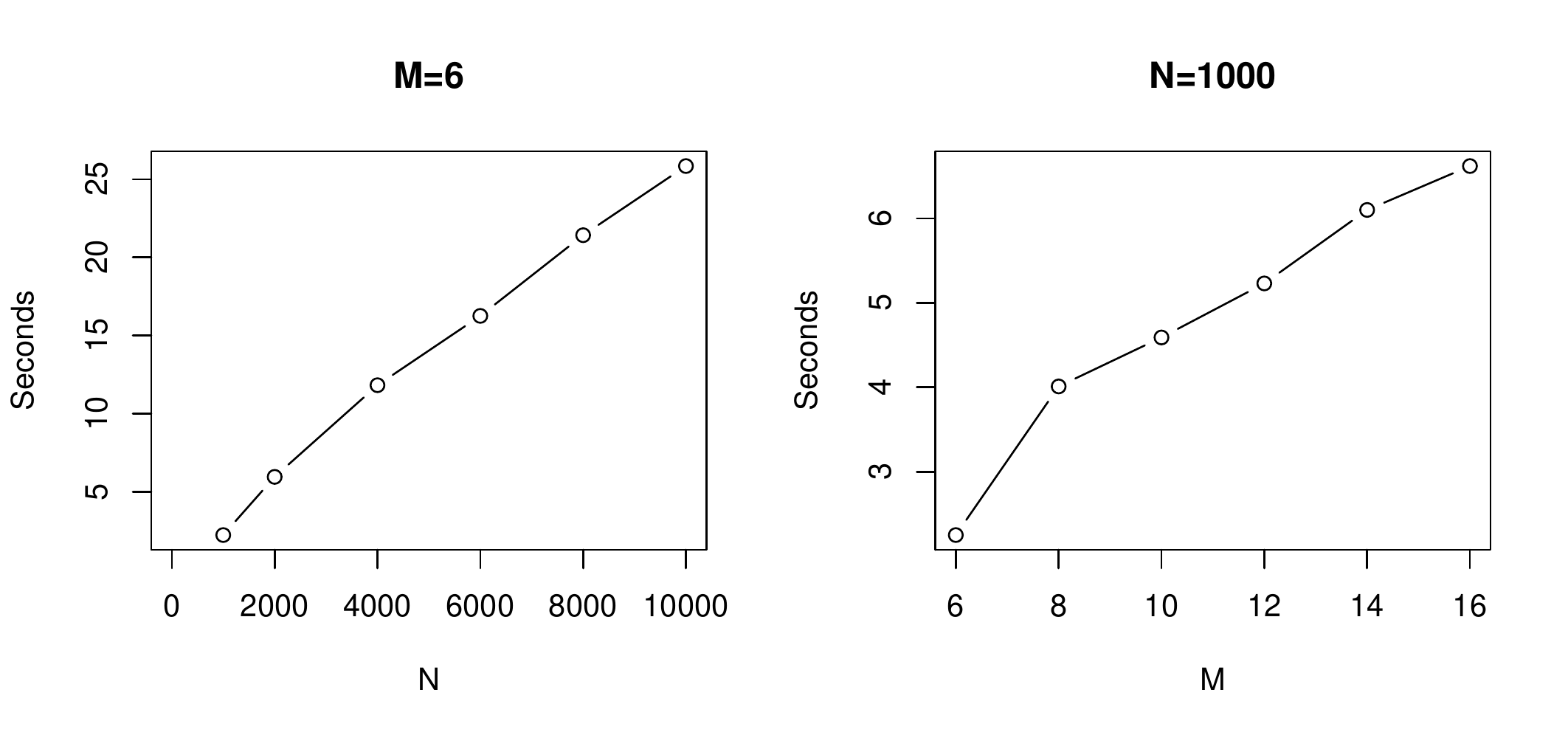}
    \caption{Computation time of CPMs with independence working correlation with large sample sizes (1000 to 10000) and large cluster sizes (6 to 16).}
    \label{fig:computation_time_ind}
\end{figure}

We further demonstrate the computation efficiency of CPMs with independence working correlation by running simulations with large sample sizes and cluster sizes. Results are shown in Figure \ref{fig:computation_time_ind}.

\section{Web Appendix E}

\begin{figure}
    \centering
    \includegraphics[scale=0.4]{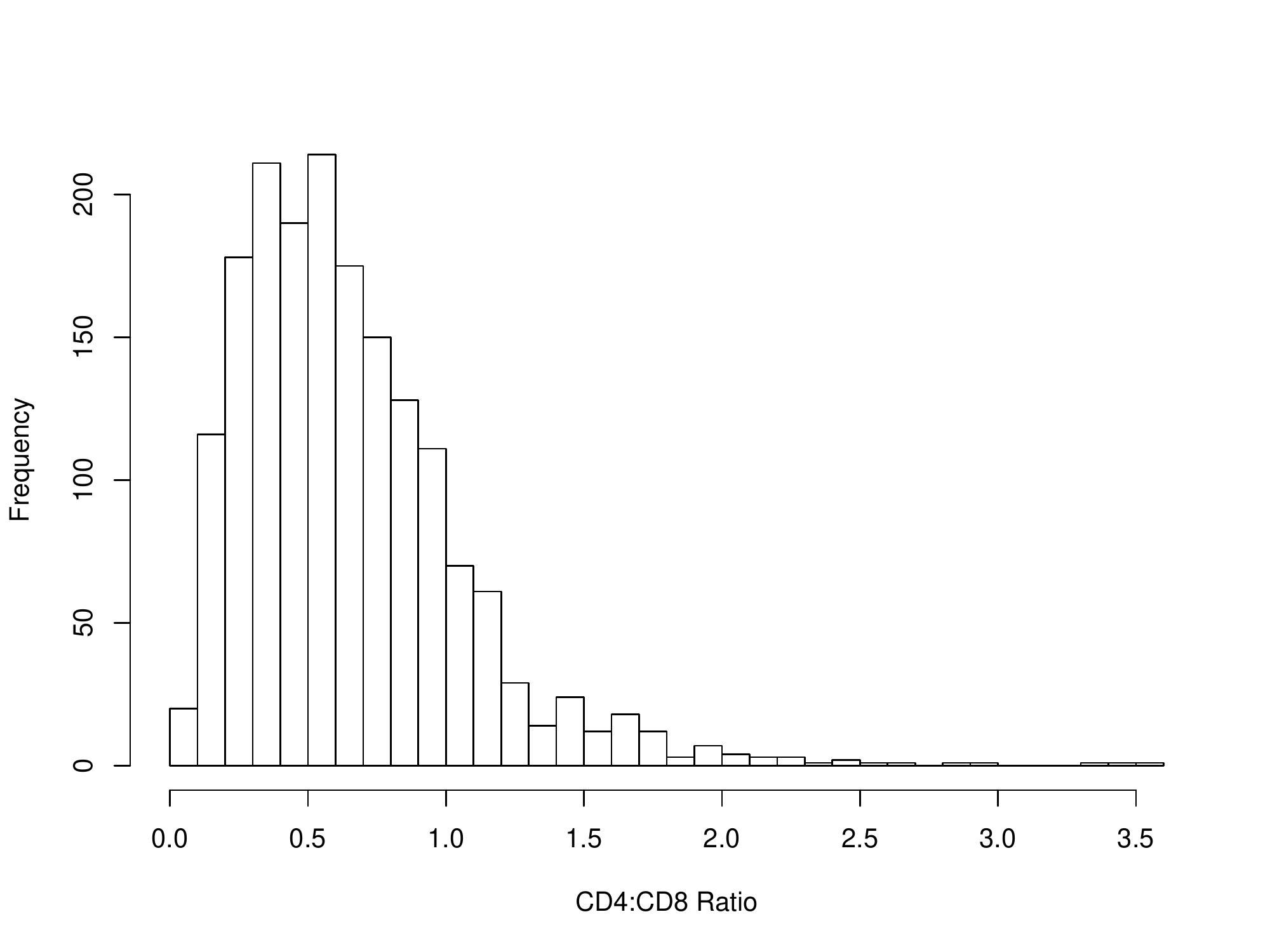}
    \caption{Histogram of CD4:CD8 ratio measured at first follow-up visit for people living with HIV and on antiretroviral therapy for a year with a suppressed viral load at the Vanderbilt Comprehensive Care Clinic (VCCC) between 1998 and 2012.}
    \label{fig:cd4cd8_first_visit}
\end{figure}

The distribution of CD4:CD8 ratio at the first follow-up visit is shown in Figure \ref{fig:cd4cd8_first_visit}.

%We analyzed the data using standard GEE methods with exchangeable working correlation after log-transforming the CD4:CD8 ratios. Results in Table \ref{tab:cd4cd8_gee} suggested that time, baseline age, sex, route, and HCV have significant association with the outcome. The conclusion and estimations were different from the results from CPM methods. Log-transformation might be a natural choice for such right-skewed responses,  but not the optimal transformation for CD4:CD8 ratios. 

\section{Web Appendix F}
\begin{figure}
    \centering
    \includegraphics[scale=0.4]{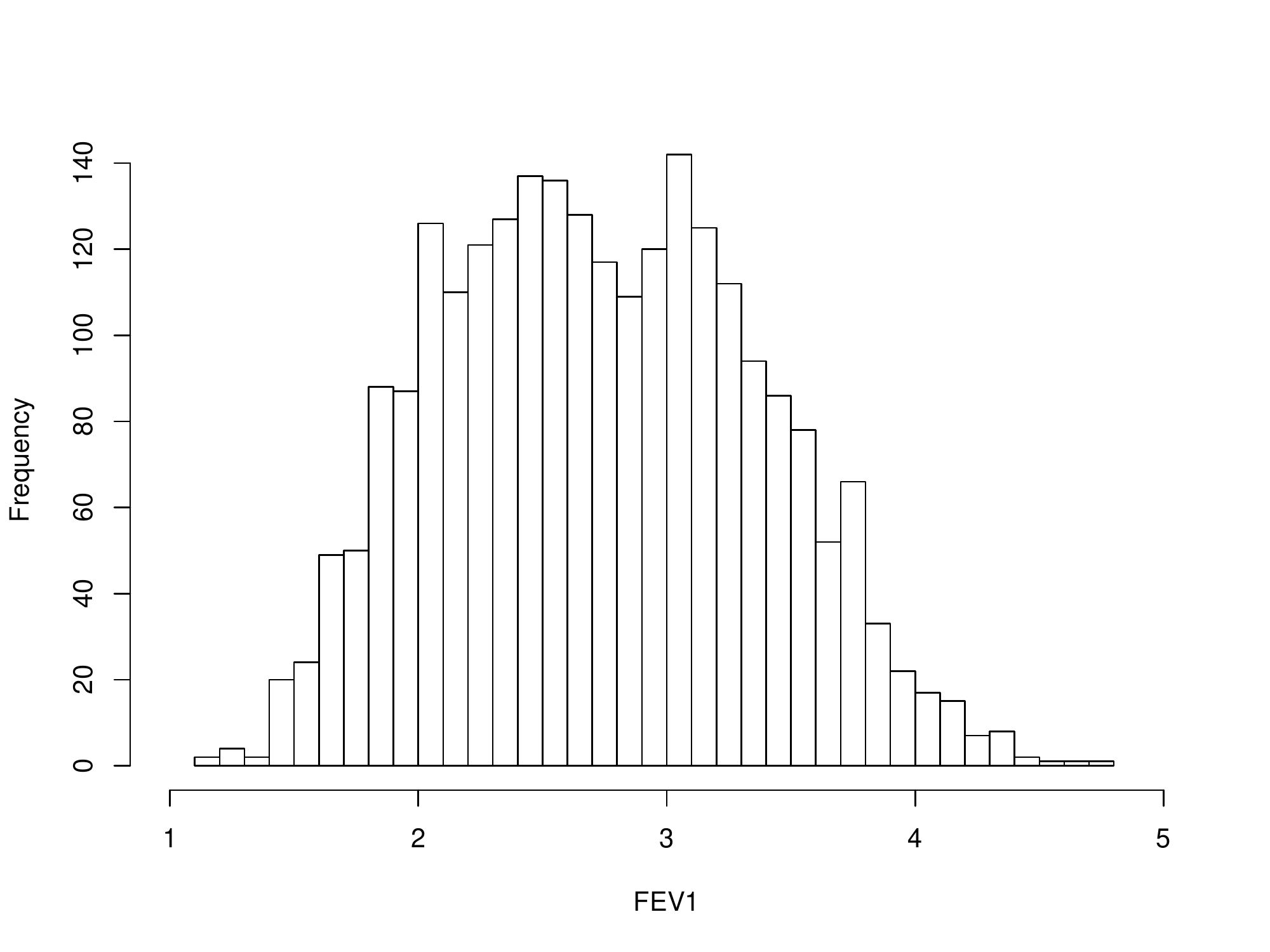}
    \caption{The histogram of the FEV1 measured at the first follow-up visit of participants in The Lung Health Study who were smokers for all 5 visits with at minimum 2 visits.}
    \label{fig:lung_first_visit}
\end{figure}

The distribution of FEV1 at the first follow-up visit is shown in Figure \ref{fig:lung_first_visit}.

\begin{figure}
    \centering
    \includegraphics[scale=0.7]{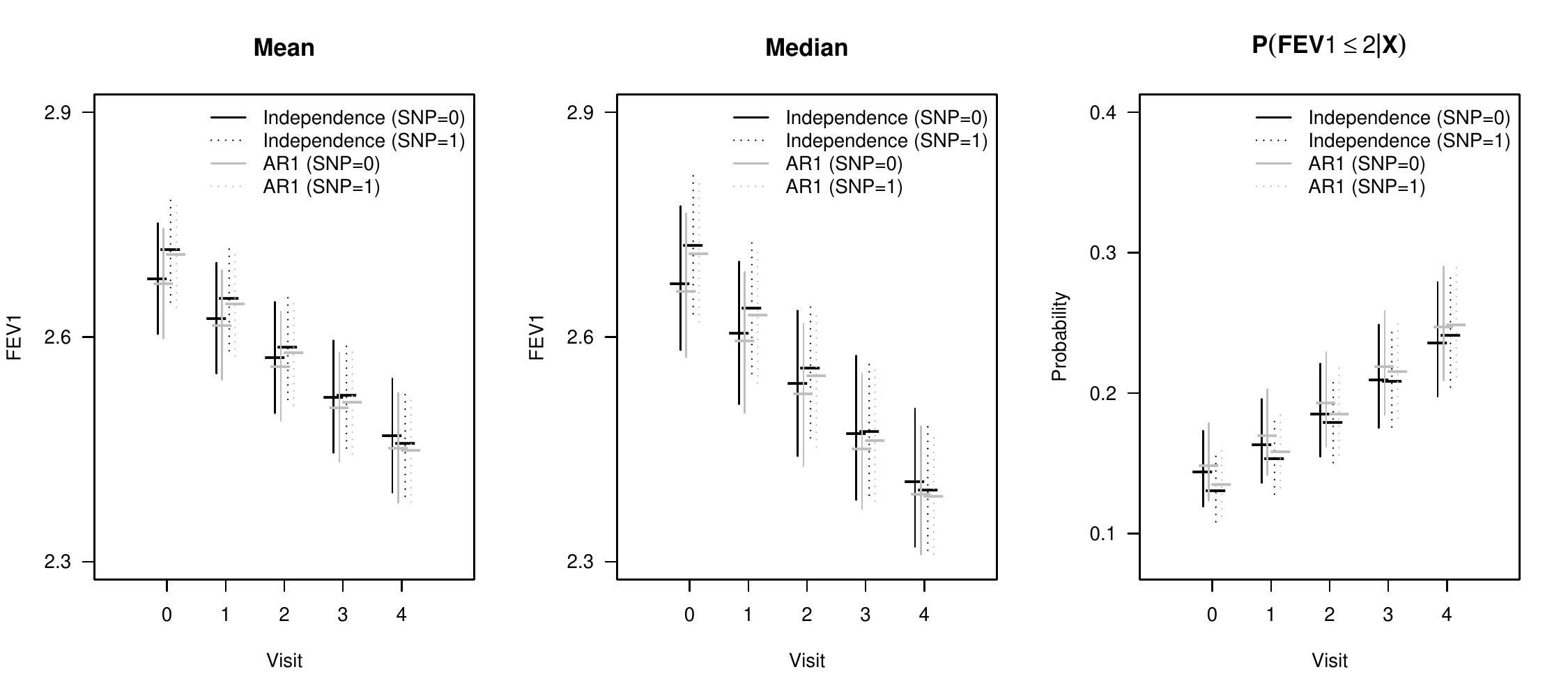}
    \caption{The estimated conditional mean FEV1, median FEV1 and the conditional probability that FEV1 is less than or equal to 2 as functions of study visit while fixing other covariates at their medians (for continuous covariates) or modes (for categorical covariates) under the circumstances that rs12194741 is present (dotted lines) and is not present (solid lines).}
    \label{fig:lung_conditional}
\end{figure}

In Figure \ref{fig:lung_conditional}, we show the conditional mean and median of FEV1, and the conditional probability of FEV1 being less than or equal to 2L  while fixing other covariates at median (continuous covariates) or mode (categorical covariates).

\newpage

\bibliography{ref.bib}